\documentclass[acmsmall,screen]{acmart}

\acmConference[POPL 2025]{52nd ACM SIGPLAN Symposium on Principles of Programming Languages}{January 2025}{Denver, United States}
\renewcommand\footnotetextcopyrightpermission[1]{} 
\AtBeginDocument{%
  }

\usepackage{booktabs}
\usepackage{filecontents}

\usepackage{graphicx}
\usepackage[many]{tcolorbox}

\usepackage{mathtools,amssymb,latexsym,amsfonts,stmaryrd}
\usepackage{bbm}
\usepackage{pbox}
\usepackage{xfrac}
\usepackage{booktabs}
\usepackage{xcolor}
\usepackage{threeparttable}
\usepackage{amsthm}
\usepackage{float}
\usepackage{multirow}
\usepackage{tikz}
\usepackage{mathrsfs}
\usepackage{mathpartir}
\usepackage{subcaption}
\usepackage{algorithm}
\usepackage{algorithmicx}
\usepackage{url}
\usepackage{syntax}
\usepackage{framed}
\usepackage[noend]{algpseudocode}
\usepackage{flushend}
\usepackage{listings}
\usepackage{moresize}
\usepackage{wrapfig}
 
\usepackage{seqsplit}
\usepackage{alltt}
\usepackage{xspace}
\usepackage{enumitem}

\newcommand{\revise}[2]{{\color{red}{\ifx&#1&\else- #1\fi}} {\color{ForestGreen}{\ifx&#2&\else+ #2\fi}}}%
\renewcommand{\revise}[2]{#2}%

\usetikzlibrary{arrows,patterns, decorations.pathreplacing}

\newcommand{\F}{Fig.}

\newcommand{\T}{Table}
\renewcommand{\S}{Sec.}

\newcommand{\ignore}[1]{}

\newcommand{\tools}{\textsc{Dsel}}
\newcommand{\toolg}{\textsc{Dgen}}
\newcommand{\bench}{\textsc{Bench}}
\newcommand{\myframe}{\textsc{DataScope}\xspace}
\usepackage{pifont}

\usepackage{amsmath, listings, amsthm, amssymb, proof, xspace}
\usepackage{bbding}

\newcommand{\finding}[2]{
  \smallskip
  \smallskip
\begin{tcolorbox}[width=\linewidth,boxrule=0pt,top=1pt, bottom=1pt, left=1pt,right=1pt, colback=gray!20,colframe=gray!20]
\textbf{Implication #1:} 
{#2}
\end{tcolorbox}}

\newcommand{\parh}[1]{\noindent\textbf{#1}}

\begin{document}

\title{API-guided Dataset Synthesis to Finetune Large Code Models}

\author{Zongjie Li}
\affiliation{%
 \institution{The Hong Kong University of Science and Technology}
 \country{Hong Kong SAR}}
\email{zligo@cse.ust.hk}

\author{Daoyuan Wu}
\affiliation{%
 \institution{The Hong Kong University of Science and Technology}
 \country{Hong Kong SAR}}
\email{daoyuan@cse.ust.hk}
\authornotemark[1]

\author{Shuai Wang}
\authornote{Corresponding authors.}
\affiliation{%
 \institution{The Hong Kong University of Science and Technology}
 \country{Hong Kong SAR}}
\email{shuaiw@cse.ust.hk}

\author{Zhendong Su}
\affiliation{%
 \institution{ETH Zurich}
 \country{Switzerland}}
\email{zhendong.su@inf.ethz.ch}

\begin{abstract}
    Large code models (LCMs), pre-trained on vast code corpora, have demonstrated remarkable performance across a wide array of code-related tasks. Supervised fine-tuning (SFT) plays a vital role in aligning these models with specific requirements and enhancing their performance in particular domains. However, synthesizing high-quality SFT datasets poses a significant challenge due to the uneven quality of datasets and the scarcity of domain-specific datasets.

    Inspired by APIs as high-level abstractions of code that encapsulate rich
    semantic information in a concise structure, we propose \myframe, an
    API-guided dataset synthesis framework designed to enhance the SFT process
    for LCMs in both general and domain-specific scenarios. \myframe\ comprises
    two main components: \tools\ and \toolg. On one hand, \tools\ employs API
    coverage as a core metric, enabling efficient dataset synthesis in general
    scenarios by selecting subsets of existing (uneven-quality) datasets with
    higher API coverage. On the other hand, \toolg\ recasts domain
    dataset synthesis as a process of using API-specified high-level
    functionality and deliberately-constituted code skeletons to synthesize
    concrete code.

    Extensive experiments demonstrate \myframe's effectiveness, with models
    fine-tuned on its synthesized datasets outperforming those tuned on
    unoptimized datasets five times larger. Furthermore, a series of analyses on
    model internals, relevant hyperparameters, and case studies provide
    additional evidence for the efficacy of our proposed methods. These findings
    underscore the significance of dataset quality in SFT and advance the field
    of LCMs by providing an efficient, cost-effective framework for constructing
    high-quality datasets. This contribution enhances performance across both
    general and domain-specific scenarios, paving the way for more powerful and
    tailored LCMs.

\end{abstract}

\maketitle

\section{Introduction}
\label{sec:introduction}

\begin{figure}[!t]
    \centering
    \includegraphics[width=1.0\linewidth]{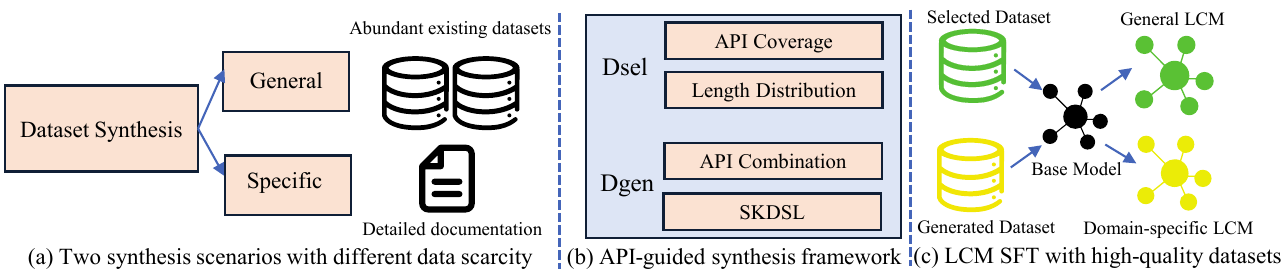}
    \caption{Overview of the proposed API-guided dataset synthesis framework,
    from its (a) targeted scenarios to (b) methodology, and further to (c) usage
    for enhancing LCM SFT in general and domain-specific scenarios.}
    \label{fig:overview-whole}
\end{figure}

Large language models (LLMs) have demonstrated remarkable performance across a wide range of tasks following extensive pre-training~\cite{kojima2022large,thapa2023humans}. In the domain of code-related tasks, large code models (LCMs) such as CodeLlama~\cite{codellama} and StarCoder~\cite{li2023starcoder} have exhibited impressive capabilities in program understanding and generation, supporting various real-world applications. However, despite their vast knowledge acquired through training on enormous datasets, these base models may not achieve optimal performance across all use cases out-of-the-box. As illustrated in \F~\ref{fig:overview-whole}(a) and (c), to further align models with diverse requirements—including enhancing general code generation capabilities~\cite{codellama,StarCoder} or specializing in specific codebases or domains (supporting commercial products like deep learning~\cite{Pecan} or security-related~\cite{secpalm} assists)—researchers often employ additional datasets to fine-tune base models, yielding more powerful and customized LCMs.

Among the various fine-tuning techniques proposed, supervised fine-tuning (SFT) has emerged as a critical approach for enhancing LLM capabilities. SFT leverages the knowledge acquired during pre-training while aligning models with human expectations~\cite{wei2021finetuned,chung2024scaling,brown2020language}. This process involves further training the models on carefully curated instruction datasets, typically comprising formatted instruction-response pairs. These pairs, represented as (INSTRUCTION, RESPONSE), consist of human-provided tasks or queries (INSTRUCTION) and the corresponding desired outputs (RESPONSE) that the model should generate~\cite{zhou2024lima,cao2023instruction,dong2023abilities,ouyang2022training}.

Given the importance of high-quality SFT datasets for LCMs, various approaches
have been developed to create and curate such datasets. These methods include
the collection of real-world code snippets~\cite{weimagicoder} and the use of
programming concepts and keywords (e.g., recursion and loops) to guide LLMs in
dataset construction~\cite{WizardCoder}. These efforts have resulted in the
generation of extensive datasets, as exemplified by Nemotron-4's 800k
examples~\cite{adler2024nemotron}. While such large datasets offer potential
benefits, they also present practical challenges in the SFT process. Notably,
the computational resources required for processing extensive datasets can be
substantial, which is particularly relevant given recent findings demonstrating
that comparable performance can be achieved with as few as 2k high-quality
examples~\cite{zhou2024lima}. Additionally, the generated data often focuses on
solving problems using basic Python operations, potentially limiting its
adaptability and efficacy in domain-specific scenarios. Consequently, as
illustrated in \F~\ref{fig:overview-whole}(a), researchers currently face a
dichotomy: an \textit{overabundance} of datasets with uneven quality for
general scenarios, and a \textit{scarcity} of high-quality, domain-specific
datasets for specialized applications.

To address these challenges, we propose \myframe, an API-guided dataset
synthesis framework designed to enhance the fine-tuning performance of LCMs.
\myframe\ offers a comprehensive solution for dataset synthesis in both general
and domain-specific scenarios, distinguishing itself from previous approaches.
On one hand, it aims to ``distill'' existing overly abundant SFT datasets to
form concise, high-quality SFT datasets for general scenarios, thereby improving fine-tuning performance while simultaneously reducing the cost of the SFT process.
On the other hand, \myframe\ supports the automated generation of
domain-specific data to facilitate SFT in specialized contexts (e.g., scientific computing and data visualization), enabling the efficient construction of high-quality SFT datasets without
requiring real-world data or being constrained by particular powerful, proprietary
LLMs.

The key observation underpinning \myframe\ is the significant benefit APIs
provide to LCMs. APIs, as high-level abstractions of code, encapsulate abundant
semantic information within a \textit{concise token structure}. As will be shown
in~\S~\ref{sec:motivation-examples}, our key insight is that such API-offered
conciseness facilitates LCMs' comprehension, alters their internal behavior
towards more accurate interpretation, and, consequently, improves their
performance. Moreover, the power of APIs can further benefit the two scenarios
in LCM SFTs: in contexts with overly abundant existing datasets, API coverage
can serve as a core metric for \textit{subset} selection, enabling LCMs to achieve
superior performance without learning from too many examples. Conversely, in
data-scarce environments, combining diverse APIs provides an effective mechanism
for constructing domain-specific datasets from scratch.

\myframe\ concretizes the above insights and observations into two key
components: \tools\ and \toolg, designed to address dataset synthesis in both
general and domain-specific scenarios. As illustrated
in~\F~\ref{fig:overview-whole}(b), \tools\ employs API coverage as its primary
metric, efficiently selecting subsets from existing datasets while also taking
into account the length distribution of the code examples to ensure diversity in
complexity. \toolg, on the other hand, leverages API information to control the
generated program functionality, utilizing a domain-specific language we
designed in~\S~\ref{subsubsec:instruction-generation}, called Skeleton
Domain-Specific Language (SKDSL)
to provide code snippet skeletons that guide the code structure, all without
requiring real-world data.

Our experimental results not only demonstrate the effectiveness of \myframe, but
also provide a series of valuable insights into API-guided dataset synthesis for
LCM fine-tuning. \tools\ efficiently selects subsets with high API coverage from
existing datasets, enabling models of varying sizes to achieve performance
comparable to or exceeding that of models fine-tuned on entire datasets. This
performance improvement scales with model size. Notably, fine-tuning a 34
billion parameter LCM on just 5\% of the data selected by \tools\ achieves 110\% of the performance compared to fine-tuning on the full dataset. Beyond
selection, \toolg\ conceptualizes dataset generation as a process of transforming high-level requirements into concrete implementations.
It not only facilitates
domain-specific dataset synthesis without relying on real-world code snippets as
seeds but also decomposes complex tasks into more manageable components through
API combination and SKDSL utilization. Furthermore, this design allows for
querying LLMs with limited
capabilities to construct domain-specific programs, enabling the automated
generation of high-quality data at reduced cost and without additional data or
human effort. 
Specifically, domain-specific SFT datasets of 4,000 examples can be generated at a cost of only 3 USD, whereas a similar scale and quality human-written dataset require over 334 volunteers~\cite{kopf2024openassistant}.
This capability further enhances the 
SFT process in developing powerful models, offering a cost-effective and efficient pathway to domain-specific model enhancement.
In summary, our contributions are as follows:
\begin{itemize}
    \item 
    Drawing inspirations from how APIs facilitate low-cost and effective LCM
    code comprehension, we propose \myframe, a simple yet highly effective
    API-guided program dataset synthesis framework to enhance LCM fine-tuning
    process from two key usage scenarios --- general and domain-specific usages.

    \item For general SFT scenarios, we introduce \tools, which employs a greedy
    algorithm to efficiently select high-quality subsets from existing datasets.
    We propose well-designed, API-based objectives in the selection process,
    enabling LCMs to achieve superior performance with significantly less data.

    \item For domain specific SFT scenarios, we present \toolg, which automates
    the generation of high-quality, tailored datasets. \toolg\ takes into
    account diverse APIs to achieve comprehensive coverage and uses SKDSL to
    provide code snippet skeletons, streamlining the creation process without
    relying on real-world data or proprietary LLMs.

    \item We demonstrate \myframe's effectiveness through comprehensive
    experiments, analyzing enhanced LCMs' performance on relevant benchmarks and
    their internal behavior in both general and domain-specific scenarios. Our
    multi-faceted evaluation, including human assessment, validates our method's
    efficacy and provides insights into the impact of API-guided dataset
    synthesis on LCM SFT.
\end{itemize}

\section{Background}
\label{sec:background}

This section introduces the background of LCMs and SFT. We explore the evolution
of LCMs and clarify two distinct scenarios in LCM fine-tuning: general and
domain-specific.

\subsection{Large Code Models}
The rapid advancements in deep learning (DL) techniques have led to the development of DL models capable of generating code with near-human or even superhuman performance~\cite{wei2021finetuned,chung2024scaling,brown2020language}. These models are often integrated into development workflows through APIs (e.g., GPT-4~\cite{gpt4}) or IDE plugins (e.g., Codex~\cite{Codex}), revolutionizing the way developers write code by improving both efficiency and quality.
The majority of state-of-the-art LLMs employ the Transformer~\cite{vaswani2017attention} architecture, which relies on attention mechanisms to enable tokens to communicate and exchange information with each other. This architecture allows LLMs to generate text sequentially, predicting the next token based on the preceding context~\cite{radford2018improving}. Building upon the success of LLMs, LCMs have emerged as a specialized variant tailored specifically for code-related tasks, leveraging the same underlying architecture.
The development of a well-performed LCM typically involves a two-step process. First, a foundation model is selected and further pre-trained on a vast corpus of code, resulting in a ``base model.'' Second, the base model undergoes fine-tuning on a task-specific dataset using various fine-tuning techniques, ultimately yielding a fine-tuned model optimized for the desired task~\cite{brown2020language}. A notable example is the CodeLlama base model, which is derived from the Llama 2 foundation model and offers instructed version models~\cite{codellama}.

\subsection{Supervised Fine-tuning}
Formally, the SFT process of the target model can be outlined as follows: for the specific domain $d$ with context $c^d$, each task example $(x^d, y^d)$ is utilized to update the model parameters. This update aims at minimizing the loss function that measures the disparity between the data distribution and the target model distribution, as expressed below:
\begin{equation}
L_{\text{SFT}}(\theta) = -\log f_\theta(y^d | c^d, x^d),
\end{equation}
Overall, this function seeks to minimize the negative log-likelihood of the
target output $y^d$ given the context $c^d$ and input $x^d$, with respect to the
model parameters $\theta$. $L_{\text{SFT}}$ converges when the generated
response $\hat{y}$ matches $y^d$, i.e., the distribution of fine-tuned model
aligns with the task dataset distribution. Compared to other fine-tuning methods
such as Reinforcement Learning from Human Feedback
(RLHF)~\cite{ouyang2022training} or Direct Preference Optimization
(DPO)~\cite{rafailov2024direct}, SFT is more efficient and effective, as it does
not require a human preference dataset. Consequently, SFT becomes a standard
procedure for developing high-quality general-purpose
LLMs~\cite{bai2023qwen,ouyang2022training} and has proven invaluable for
customizing these models across numerous domains, such as
medicine~\cite{singhal2023large}, finance~\cite{cheng2024adapting}, and various
other fields, significantly enhancing their applicability and effectiveness in
specialized contexts.

Notably, SFT methods can be further categorized into two main approaches: (1)
full parameter supervised fine-tuning (SFT) and (2) parameter-efficient
fine-tuning (PEFT). Although PEFT demonstrates high performance while using
fewer parameters, studies~\cite{ghosh2024closer,zhang2024scaling} have shown
that it primarily assists the model with response initiation and extracts most
of the response from pre-trained knowledge. In other words, PEFT does not
significantly contribute to the model's ability to acquire new knowledge.
Therefore, in this study, we focus on the full parameter fine-tuning approach
and refer to it as the SFT.

\subsection{Two Mainstream LCM Fine-tuning Scenarios}

The primary purpose of fine-tuning LCMs is to enhance their performance on code
generation tasks and align them with human instructions. Based on whether the
fine-tuned models are intended for specific domains, LCM fine-tuning can be
categorized into two mainstream scenarios, general and domain-specific, each
facing distinct data scarcity, as illustrated in \F~\ref{fig:overview-whole}(a).

\parh{General Scenario.} In this scenario, the fine-tuned model is designed as a
universal code generation tool aimed at improving its general generation
capabilities. A prime example is the instructed version of
CodeLlama~\cite{codellama}, which demonstrates superior performance across a
wide range of code generation tasks compared to its base model. Researchers and practitioners working on fine-tuning general LCMs often face an abundance of existing datasets, such
as OSS-instruct~\cite{weimagicoder} and CodeExercise~\cite{CodeExercise}, which
are typically derived from various online code repositories and programming
forums using LLMs.

While the generation methods behind these datasets attempt to synthesize a
diverse range of code snippets, they are inherently constrained by the quality
of online code samples and the restricted number of programming forums.
Consequently, these datasets often suffer from a paradox of quantity over
quality, containing a large volume of data with inconsistent quality. For
researchers aiming to construct SFT datasets to enhance a model's general
performance, an efficient and cost-effective dataset synthesis approach would
involve creating new SFT datasets by judiciously selecting from existing
datasets.

\parh{Domain Specific Scenario.} In real-world applications, LCMs are often
designed to meet the unique requirements of their target audience. For example,
Pecan~\cite{Pecan} focuses on generating machine learning code, while
SQLCoder2~\cite{Sqlcoder2} specializes in generating SQL-related code. In such
cases, fine-tuning aims to enhance the model's performance in specific domains
by leveraging domain-specific datasets and integrating pertinent domain
knowledge into the model.

Users aiming to tune domain LCMs frequently encounter a key challenge: the
scarcity of high-quality, domain-specific datasets. Curating such datasets
demands expertise in the target domain and often involves manual collection,
cleaning, and annotation of code samples—a process that is both time-consuming
and resource-intensive~\cite{zhou2024lima,daniel2018quality}. Moreover, many
valuable datasets may originate from proprietary company codebases, limiting
their public availability and further exacerbating the data scarcity issue.
Given these constraints, there is a pressing need for an efficient and adaptable framework that facilitates the synthesis of high-quality, domain-specific datasets. Such a framework would ideally minimize the reliance on extensive manual data generation while still capturing the nuances and complexities of the target domain, thereby enabling more effective and targeted fine-tuning of LCMs for specialized applications.

\section{A Motivation Example}
\label{sec:motivation-examples}

While APIs have been widely recognized for their effectiveness in program
synthesis~\cite{sotiropoulos2024api}, this paper presents a novel perspective on
their role in enhancing the performance of LCMs through SFT dataset synthesis.
To ease understanding, this section provides a motivating example to
illustrate the benefits of using APIs in LCM program comprehension. Our example
aims to elucidate how API-level abstractions can significantly improve an LCM's
ability to understand and process code semantics, thereby underpinning our
approach to SFT dataset synthesis.

\begin{figure}[!t]
    \centering
    \includegraphics[width=1.0\linewidth]{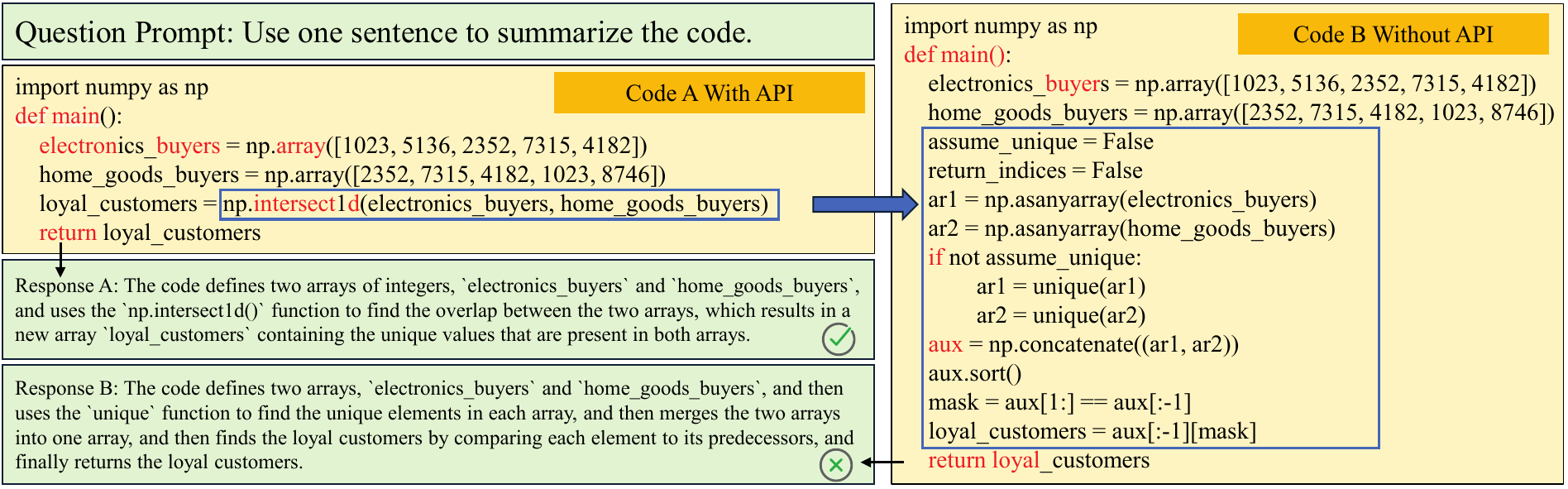}
    \caption{An illustrative example showing the crucial role of APIs in helping
    LCMs understand programs. For readability, we have shortened the example
    code and simplified the API calls. The blue box on the left shows the
    original highly abstract code using APIs, while the right one displays the
    code after we manually expanded the API calls. The tokens highlighted
    \textcolor{red}{in red} are the top 20 tokens with the highest average
    attention scores across all attention heads in the model's last layer.}
    \label{fig:API-example}
    \vspace{-15pt}
\end{figure}

Considering \F~\ref{fig:API-example}, where we present two semantically
equivalent code snippets, both aim to find customers who purchase both
electronics and home goods (loyal_customers). Code snippet A utilizes the highly
abstracted \texttt{np.intersect1d()} API call, while code snippet B replaces the
API call with its official implementation. We input these code snippets into the
llama2-7B-hf model, prompting it to summarize the program functionality. The
llama model yields responses A and B, respectively (as shown in the bottom left
of \F~\ref{fig:API-example}).

Moreover, to gain insight into the model's internal processing, we analyze the
attention mechanisms within the LCM. Attention scores, a key component of
Transformer-based models, reflect the model's understanding of input
importance~\cite{vaswani2017attention}. We calculate the average attention score
for each token across all attention heads in the model's final layer. The top 20
tokens with the highest average attention scores are highlighted in red for both
programs. This visualization offers a glimpse into how the LCM prioritizes different
parts of the input, potentially revealing differences in its comprehension of
highly abstracted API calls versus their expanded implementations.
Based on the example, we summarize the following three key advantages of using
API abstractions in LCM program comprehension:

\begin{enumerate}[label=\textbf{\arabic*}.]
    \item \textbf{Token Efficiency:} The most apparent advantage of using APIs
    is the substantial reduction in token count. The API-abstracted code
    (snippet A) is significantly shorter than its expanded counterpart (snippet
    B), comprising only 49\% of the original length. Importantly, it has been
    pointed out that the computational intensity and resource requirements of
    LCM fine-tuning and inference processes correlate positively with input
    token count~\cite{vaswani2017attention, han2024parameter}. Therefore, the reduced token count in API-abstracted
    code snippets implies that the LCM's resource consumption will be
    significantly lower, and we envision that API abstraction will support more
    efficient LCM fine-tuning and inference.
    
    \item \textbf{Semantic Comprehension:} A closer examination reveals the
    benefits of APIs in enhancing LCMs' understanding of program semantics.
    Real-world API names often encapsulate program functionality, enabling more
    concise and semantically rich code. Analyzing the model's responses, we
    observe that highly abstracted APIs more accurately reflect the program's
    functionality. Response A precisely captures the program's aim of finding
    customers who purchased both electronics and home goods, while response B
    merely describes the function's operations. This demonstrates how API
    abstraction swiftly mitigates LCMs' hurdles in understanding complex
    program semantics, thereby enhancing their performance.

    \item \textbf{Internal Attention Patterns:} When examining the LLM's
    internal behavior, we notice that API calls typically receive higher
    attention scores. In contrast, manually expanded programs fail to allocate
    similar levels of focus on key tokens. This analysis further suggests that
    the high-quality language abstraction provided by APIs augments the model's
    comprehension ability towards crucial semantic elements.
\end{enumerate}

Overall, our key insight is that APIs unleash a new dimension in SFT dataset
synthesis by augmenting the comprehension capabilities and reducing the
resource consumption of LCMs with code API-level abstractions. Given the
pervasive use of APIs in modern programming, this insight is particularly
valuable, as the extra cost of involving APIs is often negligible, while the
benefits are substantial. This insight is the cornerstone of our API-guided
framework, which consists of two components to achieve well-performing and
low-cost SFT dataset synthesis under different scenarios. The subsequent
sections will introduce these two components in detail.

\section{API-guided Dataset Selection (\tools): Design and Evaluation}
\label{sec:tools}

We first define the problem in \S~\ref{subsec:problem-formulation}: we aim at
selecting a subset from an SFT dataset to optimize the performance of a model
trained on this subset.
In the following subsections, we first introduce \tools, and present the
implementation and evaluation results in the following subsections.

\subsection{Problem Formulation}
\label{subsec:problem-formulation}

Given an SFT dataset $D = {(x_i, y_i)}_{i=1}^N$, where each example consists of
an instruction $x_i$ and its corresponding code $y_i$, our goal is to select a
subset $D' \subseteq D$ of size $n$ ($n \leq N$) such that a model trained on
this subset achieves maximum performance on general code generation tasks, which
can be formulated as:

\begin{equation}
\label{eq:basic-problem}
\begin{aligned}
\max_{D' \subseteq D} \quad & \text{Performance}(D') \
\quad \textrm{s.t.} \quad & |D'| = n.
\end{aligned}
\end{equation}

Despite the rather straightforward formulation, predicting the performance of a
fine-tuned model is challenging. Drawing inspiration from our key observations
in~\S\ref{sec:motivation-examples}, which demonstrated the influence of APIs on
LCM code comprehension, we propose using API coverage as a proxy measure for
performance. Furthermore, building on previous research that highlights the
importance of diversity in dataset
quality~\cite{zhou2024lima,adler2024nemotron,peng2023instruction}, we also
consider the diversity of code lengths in the selected subset. To quantify this
aspect, we introduce a length diversity measure, $\text{LenDist}(D', D)$, which
assesses the similarity between the length distributions of the subset $D'$ and
the original dataset $D$. Incorporating these design considerations, we
reformulate our optimization problem as:

\begin{equation}
\begin{aligned}
\max_{D' \subseteq D} \quad & \text{APICoverage}(D') \
\quad \textrm{s.t.} & |D'| = n, \
& \text{LenDist}(D', D) \leq \tau,
\end{aligned}
\end{equation}

\noindent where $\text{APICoverage}(D')$ represents the number of unique APIs
covered in subset $D'$, and $\tau$ is a length diversity threshold. This
formulation aims to maximize API coverage while maintaining a representative
distribution of code lengths.

\parh{SFT Dataset Selection vs. Test Case Selection.}
Test case selection is a well-studied problem in software testing, with the
objective to select a subset of test cases $T' \subseteq T$ from a large pool
$T$ while maintaining software quality and reducing the number of test
cases~\cite{yoo2007pareto, lou2015mutation}.
Although both test case selection and dataset selection can be formulated as
optimization problems, there is a notable difference: as new, non-trivial
examples are continuously added to the existing test case set, code coverage
will monotonically increase, such that the software will be tested more
thoroughly. In contrast, such a monotonicity does \textit{not} hold for SFT;
adding more code snippets does not necessarily lead to better model performance
on the task, as it may introduce more noise or cause
overfitting~\cite{glorot2010understanding}, resulting in \textit{performance
degradation}.

\subsection{Selection Algorithm}
\label{subsec:tools-methodology}

Although the optimization problem here can be solved through an exhaustive
search approach, it quickly becomes infeasible for large datasets due to the
combinatorial explosion of possible subsets. Therefore, we propose a sub-optimal
greedy algorithm in \tools\ to efficiently solve this problem. The main idea is
to iteratively select examples that maximize the incremental API coverage while
maintaining the diversity of code lengths. The detailed steps are shown in
Algorithm~\ref{alg:tools}. We first initialize the selected indexes subset as
empty and the API set as empty (line 2). We also initialize a list
$bucket_{cnt}$ to keep track of the number of examples selected for each length
bucket and calculate its initial values by distributing the total number of
examples to be selected ($n$) across the buckets based on the length
distribution of the original dataset (lines 3-4).

In each iteration (lines 5-17), we select the example that adds the most uncovered APIs to the current API set. To ensure diversity, we prioritize examples from underrepresented length buckets. We traverse each non-empty bucket (line 7) and each example within the current bucket (line 8).
For each valid example (i.e., not already selected and belongs to the current bucket), we calculate the number of new APIs it would add to the API set (lines 9-10). We update the best example if it adds more new APIs than the current best (lines 11-12). If the best example is found for the current bucket, we break the loop and move to the next iteration (lines 13-14).

After selecting the best example, we add its index to the selected indexes subset, update the API set with its APIs, and decrement the count for its corresponding length bucket (lines 15-17). We repeat this process until $n$ examples are selected or no more examples can be added to improve the API coverage.

    \begin{algorithm}
        \caption{SFT data selection using \tools.}
        \label{alg:tools}
        \begin{algorithmic}[1]
        \small
        \Require $n$ (Number of cases to select), $api\_stats$ (API statistics for each case), $case\_lens$ (Lengths of each case), $buckets$ (Number of buckets for length distribution)
        \Ensure Indices of selected cases
        
        \Procedure{SelectTopAPI}{$n, api\_stats, case\_lens, buckets$}
            \State Initialize $selected \gets [], api\_set \gets \{\}$ \Comment{Init selected indices and API set}
            \State Initialize $bucket\_cnt \gets [0] \times buckets$ \Comment{Init bucket counts}
            \State Calculate initial $bucket\_cnt$ based on $n$, $case\_lens$ and $buckets$ \Comment{Distribute $n$ cases across buckets}
        
            \For{$i \gets 1$ to $n$}
                \State Initialize $best \gets -1, max\_new \gets -1$
        
                \For{$bkt$ in prioritized order} \Comment{Traverse each non-empty bucket}
                    \For{$id, apis$ in $api\_stats$} \Comment{Traverse each case for current bucket}
                        \If{$id$ is valid for current $bkt$ and not in $selected$}
                            \State Calculate $new\_apis$ for $id$ \Comment{Number of new APIs $id$ would add to $api\_set$}
                            \If{$new\_apis > max\_new$}
                                \State Update $best \gets id, max\_new \gets new\_apis$
                            \EndIf
                        \EndIf
                    \EndFor
                    \If{$best \neq -1$}
                        \State \textbf{break}
                    \EndIf
                \EndFor
        
                \If{$best \neq -1$}
                    \State Add $best$ to $selected$, Update $api\_set$ with APIs from $api\_stats[best]$
                    \State Decrement $bucket\_cnt$ for the corresponding $bkt$ by 1 \Comment{Update bucket count}
                \EndIf
            \EndFor
        
            \State \Return $selected$
        \EndProcedure
        \end{algorithmic}
\end{algorithm}

\subsection{Experimental Setup for General Scenario}
\label{subsec:tools-experiment}

\parh{SFT Datasets.} 
We evaluate \tools\ on two datasets: CODEE and OSS. CODEE is primarily sourced from the CodeExercise dataset~\cite{CodeExercise}, which contains programming exercises covering a wide range of Python-related topics, including basic syntax, data structures, algorithm applications, and database queries. To further enhance the dataset's diversity and quality, we supplement it with the MEIC dataset~\cite{wang2024exploring}, which follows a similar data generation process and template, providing a substantial number of high-quality instruction pairs. In contrast, OSS is derived from the MagicCoder's OSS-Instruct dataset~\cite{weimagicoder}. This dataset generates a large number of instruction pairs by querying GPT-3.5 with a combination of real-world code snippets. 
To ensure data quality and consistency, we apply a further processing step to both datasets. We select Python-related examples, remove instances failing syntax checks or exceeding length thresholds, and eliminate potentially invalid or excessively long code snippets. This filtering process helps to eliminate any potentially invalid or excessively long code snippets that may adversely affect the training process. After processing, we obtain 76,512 examples from CODEE and 37,284 examples from OSS, forming the basis for evaluating our proposed SFT subset selection method.

\parh{Models.} For our SFT experiments, we select CodeLlama, a family of LCMs based on Llama 2, as our starting point. CodeLlama has demonstrated strong performance in code-related tasks and has been widely adopted in previous works~\cite{zhang2024large,ding2024cycle,ullah2024llms}, establishing itself as a reliable benchmark model. To investigate the impact of model size on the effectiveness of SFT, we consider three different sizes of the CodeLlama base model: 7B, 13B, and 34B parameters.
As noted in~\S~\ref{sec:background}, META provides both base and instructed versions of CodeLlama. We opt for the base version to ensure that our evaluation is not influenced by the fine-tuning process used in CodeLlama's instructed version, as the technical details and SFT datasets used are not clearly disclosed~\cite{codellama}. Using the base version allows us to isolate the effects of \tools\ and maintain a fair comparison across different model sizes, without potential confounding factors introduced by the instructed version's fine-tuning process.

\parh{Benchmark.} We select Humaneval~\cite{chen2021evaluating} as our benchmark, which has become a widely accepted standard for evaluating the performance of SFT models on code generation tasks~\cite{al2024traces,sun2024neural,ding2024cycle,nguyen2024beginning}. Humaneval comprises a diverse collection of programming problems accompanied by human-written instructions and corresponding code solutions. This benchmark serves as a reliable indicator of a model's general code generation capabilities.

\parh{Metrics.} In line with prior studies \cite{wei2023magicoder, li2023starcoder, luo2023wizardcoder,abdin2024phi, codellama}, we employ the Pass@k metric to assess the accuracy of LCMs in solving programming problems. This metric determines whether LCMs can pass all unit tests within $k$ generated solutions. Formally, it can be expressed as:

\begin{equation}
\mathit{Pass@k}= \frac{\sum \limits_{i=1}^{n}\prod_{j=1}^k(\mathbbm{1}(pass_{s_i^j}))}{n},
\end{equation}

\noindent where $n$ and $k$ represent the number of problems and the number of
generated solutions, respectively. $s_i^j$ denotes the $j$-th solution for the
$i$-th problem. The function $\mathbbm{1}(x)$ returns 1 if $x$ is True and 0
otherwise, and $pass(s)$ returns True if the solution $s$ successfully passes
all unit tests. Following~\cite{wei2023magicoder, li2023starcoder,
luo2023wizardcoder}, we adopt Pass@1 as our primary metric, i.e., $k=1$.

In addition to the Pass@k metric, we employ the Jensen-Shannon (JS) divergence~\cite{fuglede2004jensen} to evaluate the similarity between the length distributions of the selected subset and the original dataset. To calculate the JS divergence, we divide the range of code lengths into equal-width bins and compute the normalized frequency counts of code lengths falling into each bin for both datasets. The JS divergence between the two resulting distributions measures their similarity, ranging from 0 to 1, with smaller values indicating higher similarity. A smaller JS divergence suggests that the selected subset better represents the code length diversity of the original dataset, helping assess the representativeness of the selected subset in terms of code length distribution.

\begin{table}[!t]
    \centering
    \caption{Hyper-parameter settings.}
    \vspace{-6pt}
    \resizebox{0.65\linewidth}{!}{
    {\begin{tabular}{l|c||c|c}
    \toprule
       Hyperparameter  & Value &  Hyperparameter  & Value\\
     \midrule
      Optimizer & AdamW~\cite{KingBa15} & Warm-up steps   & 100 \\
      Learning rate & 5e-6 & Training batch size & 64 \\
      LR scheduler& Cosine Scheduler \cite{loshchilov2017sgdr}& Validation batch size & 32 \\
        Sequence Len. & 2,048 &  Adam epsilon & 1e-8 \\
      Max. gradient norm & 1.0 & Precision & BF16\\
      \midrule
      Max Gen. Tokens & 512 & Top-P & 0.95 \\
      \bottomrule
    \end{tabular}
    }
    }
    \label{tab:param}
\end{table}

\parh{Hyperparameters.} \T~\ref{tab:param} lists the hyperparameters used in our
experiments. For \tools, we set the number of buckets to 40 and the budget
constraint $n$ to 2.5\%, 5\%, 10\%, 20\%, and 25\% of the training data size.
These settings allow us to investigate the effectiveness of our method under
different data budgets. The API coverage is calculated based on the unique APIs
in the selected subset. All experiments are conducted on a server equipped with
eight H800 GPUs, each having 80GB of graphic memory. To optimize GPU memory
usage and improve training efficiency, we employ Optimizer State Sharding (ZeRO
3) techniques from DeepSpeed \cite{rasley2020deepspeed, rajbhandari2020zero}.
For larger models such as CodeLlama 34B, we further offload the optimizer state
into CPU memory to mitigate potential out-of-CUDA memory issues.
Following~\cite{wang2024exploring}, we allocate 10\% of the training samples as
a validation set. During training, we monitor the validation loss and select the
three checkpoints closest to the lowest validation loss for inference. From
these three checkpoints, we choose the best-performing one as our final model.

\subsection{Results for \tools}
\label{subsec:tools-evaluation}

In this section, we first analyze the subset selection results of \tools\ in
terms of API coverage and JS divergence. Then, we evaluate the performance of
SFT models trained on the selected subsets using the Pass@1 metric on the
Humaneval benchmark. Finally, we provide a detailed analysis of the results and
offer insights into the effectiveness of \tools.

\subsubsection{Comparative Analysis of Subset Selection Methods}
\label{subsubsec:tools-subset-analysis}

\begin{table}[!t]
    \caption{Comparative analysis of subset selection methods on \bench. API
    coverage (higher is better) and Jensen-Shannon (JS) divergence (lower is
    better) are shown for different subset sizes using Random selection,
    Clustering-and-Retrieve (CaR), and our proposed \tools.}
    \vspace{-6pt}
    \resizebox{0.6\linewidth}{!}{
    \begin{tabular}{c|c|c|c|c|c|c|c}
    \hline
    \multirow{2}{*}{Dataset} & \multirow{2}{*}{Size} & \multicolumn{3}{c|}{API Coverage (\%)} & \multicolumn{3}{c}{JS Divergence} \\
    \cline{3-8}
    & & Random & CaR & \tools & Random & CaR & \tools \\
    \hline
    \multirow{5}{*}{OSS} & 2.5\% & 6.84 & 7.01 & 18.95 & 0.0787 & 0.0886 & 0.0763 \\
    & 5\% & 13.08 & 11.55 & 39.84 & 0.0760 & 0.0739 & 0.0734 \\
    & 10\% & 26.53 & 22.45 & 55.33 & 0.0678 & 0.0781 & 0.0677 \\
    & 20\% & 25.99 & 40.06 & 67.23 & 0.0772 & 0.0801 & 0.0659 \\
    & 25\% & 31.67 & 56.26 & 77.82 & 0.0765 & 0.0798 & 0.0546 \\
    \hline
    \multirow{5}{*}{CODEE} & 2.5\% & 6.74 & 7.76 & 23.37 & 0.0546 & 0.0560 & 0.0512 \\
    & 5\% & 11.86 & 11.70 & 36.98 & 0.0538 & 0.0539 & 0.0514 \\
    & 10\% & 19.45 & 19.09 & 56.45 & 0.0543 & 0.0550 & 0.0503 \\
    & 20\% & 32.17 & 32.21 & 89.00 & 0.0547 & 0.0539 & 0.0310 \\
    & 25\% & 38.21 & 37.97 & 100.00 & 0.0543 & 0.0540 & 0.0208 \\
    \hline
    \end{tabular}
    }
    \label{tab:api-js}
\end{table}

To comprehensively evaluate the effectiveness of \tools, we conduct experiments on two datasets (OSS and CODEE) and compare \tools\ against two baselines: random selection (Random) and a clustering-based selection (CaR)~\cite{ge2024clustering}. For the Random baseline, we randomly select three sets of examples using different random seeds, perform a preliminary study, and choose the seed that yields the best results for subsequent experiments. In the CaR baseline, we follow the same setting, where we use the Sentence-BERT model~\cite{reimers2019sentence} to embed the instruction pairs and apply K-means~\cite{kmeans1979algorithm} clustering algorithm on the feature vectors processed by PCA~\cite{abdi2010principal}. We then select the top $k$ examples from each cluster as representatives.
The results are shown in \T~\ref{tab:api-js}, from which we can make the following observations:

\parh{API Coverage Analysis.} \tools\ consistently achieves significantly higher coverage compared to both Random and CaR across all subset sizes and datasets. Notably, at the 25\% subset size, \tools\ covers 77.82\% and 100\% of the unique APIs for the OSS and CODEE datasets, respectively, outperforming the baselines by 33.85\% and 41.80\%. This demonstrates the effectiveness of \tools\ in maximizing API coverage through its iterative greedy selection approach. As the dataset size increases, the API coverage of all three methods increases. However, the convergence speed is slower on OSS compared to CODEE, with \tools\ achieving a coverage increase of 58.87\% from 2.5\% to 25\% subset size on OSS, compared to a 76.63\% increase on CODEE. This indirectly reflects the challenge of finding a well-covering subset for OSS, as it stems from complex real-world code crawled from the internet. Interestingly, all methods' API coverage values are higher than the corresponding dataset percentages. We attribute this to that different examples contain varying numbers of APIs, as well as some examples are relatively easy or share common APIs, such as max(), min(), etc. Such usage overlap leads to higher API coverage values.

\parh{JS Divergence Analysis.} \tools\ consistently maintains lower divergence compared to the baselines, especially at larger subset sizes. On average, \tools\ achieves JS divergence that is 0.0128 lower than Random and 0.0203 lower than CaR. At the 25\% subset size, \tools\ achieves JS divergence of 0.0546 and 0.0208 for OSS and CODEE, respectively. These values are 0.0219 and 0.0252 lower than Random, and 0.0252 and 0.0332 lower than CaR for OSS and CODEE, respectively. This highlights the ability of \tools\ to select subsets that closely resemble the length distribution of the original dataset. Similar to the API coverage results, we observe that the JS divergence decreases more slowly on OSS compared to CODEE, with a decrease of 0.0217 from 2.5\% to 25\% subset size on OSS, compared to a decrease of 0.0304 on CODEE. This further reflects the complexity of the data distribution and the effectiveness of \tools\ in handling it.

\subsubsection{\tools's Performance Evaluation and Analysis}
\label{subsubsec:tools-performance-evaluation}

\begin{figure}[!t]
    \centering
    \includegraphics[width=1.0\linewidth]{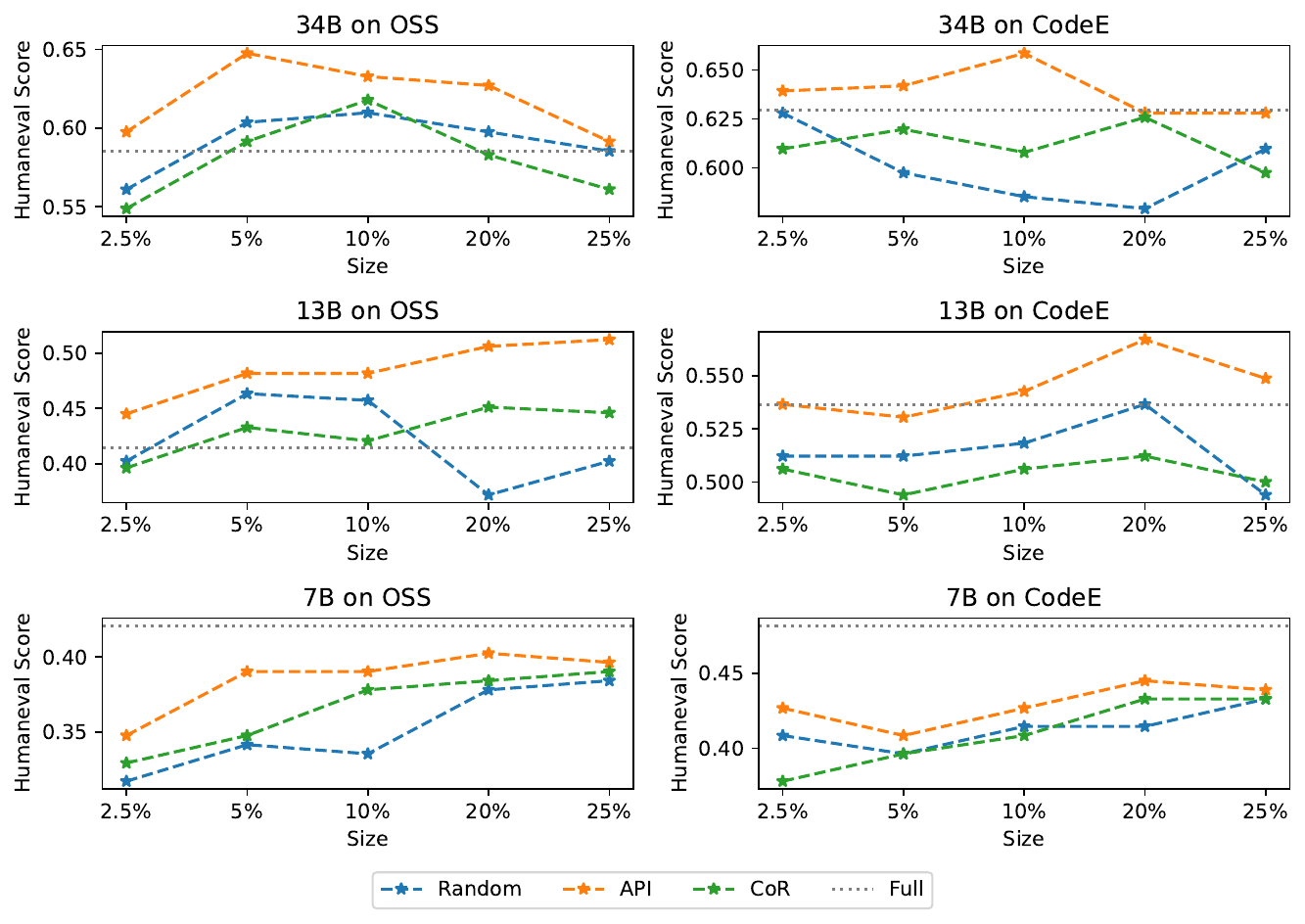}
    \vspace{-15pt}
    \caption{Pass@1 score of \tools, Random, and CAR on the Humaneval benchmark using CodeLlama models of different sizes trained on subsets of varying sizes from the OSS and CODEE datasets in \bench. }
    \label{fig:evaluation-humaneval}
    \vspace{-10pt}
\end{figure}

\F~\ref{fig:evaluation-humaneval} presents the evaluation results of \tools\ on
the Humaneval benchmark using CodeLlama models of different sizes (7B, 13B, and
34B), in which the score is represented by pass@1 rate. Overall, we view the
results as highly promising. We now analyze the results by analyzing the
``full dataset training'' and two random selection baselines.

We first investigate how models fine-tuned using \tools-selected subsets perform
compared to those trained on the full SFT dataset (denoted as ``FULL''
in~\F~\ref{fig:evaluation-humaneval}). Interestingly, we observe that for the 7B
model, using the full SFT dataset yields better results than using subsets
selected by \tools. However, this trend reverses for the 13B and 34B models,
where using fewer examples selected by \tools\ leads to better performance.
Moreover, we find that the larger the model, the smaller the number of examples
required to achieve superior performance compared to training on the full
dataset.

We hypothesize that this phenomenon can be attributed to the following reasons:
The 7B model may not have effectively learned code-related knowledge during
pretraining, thus requiring more data to assist in its learning process during
SFT. On the other hand, the 34B model has already acquired a substantial amount
of code-related knowledge during pretraining, enabling it to learn more
effectively with only a small amount of carefully selected data during SFT. This
finding has significant implications, as larger models incur higher fine-tuning
costs in terms of both time and hardware requirements. The effectiveness of
\tools\ in improving the performance of larger models while using fewer examples
demonstrates its potential to greatly reduce the computational resources needed
for SFT, making it more feasible to develop and deploy large-scale LCMs.

Besides the full dataset results, \tools\ largely improves the performance of
SFT models across all model sizes and datasets compared to the random selection
(Random) and clustering and ranking (CAR) baselines. 
On average, \tools\ outperforms Random and CAR by 7.96\% and 7.14\%. In general
code generation scenarios where syntax and semantics correctness is required,
this relative improvement means approximately 7-8\% more user queries can be
correctly processed on the first attempt. Even considering just a single round
of re-querying (i.e., simply asking again to obtain the correct result), this
improvement translates to at least 7\% reduction in computational costs, which
can be significant in large-scale applications. 

\parh{Impact of Model Size and Dataset Quality.}~First, we observe that as the model size increases from 7B to 34B, the relative improvement brought by \tools\ also increases. For instance, on the CODEE dataset, compared to the base model finetuned with two baselines, 
\tools\ achieves an average Pass@1 improvement of 4.29\% for the 7B model, while the improvement increases to 7.06\% and 5.45\% for the 13B and 34B models, respectively. This suggests that larger models generally benefit more from the intelligent selection of SFT data by \tools. Second, we notice that the performance of models finetuned on the OSS dataset is consistently lower than those tuned on the CODEE dataset by a margin of 8.29\%. This significant gap highlights the importance of dataset quality and the limitations of using open-source code to construct SFT datasets. Finally, we find that for larger models (13B and 34B), increasing the subset size does not always lead to better performance, which is consistent with the findings in \cite{dong2023abilities}.

\finding{for \tools}{Selecting subsets of existing datasets with higher API coverage not only aids in efficient dataset synthesis, but also provides increasingly significant benefits as model size scales up.}

\section{API-Guided Dataset Generation (\toolg)}
\label{sec:toolg}

Building upon our previous analysis of API impact on LCMs, both without
fine-tuning (\S~\ref{sec:motivation-examples}) and fine-tuning in general
scenarios (\S~\ref{sec:tools}), we now address the challenge of dataset scarcity
in domain usages. To complete our framework and provide a comprehensive solution
for dataset synthesis, we introduce \toolg, a component designed for generating
high-quality SFT datasets leveraging API combinations.

\subsection{Problem Statement and Approach Overview}
\label{subsec:overview-TOOLG}

Extending our research beyond the scope presented in
\S~\ref{subsec:problem-formulation}, we now confront a more complex challenge:
the absence of a pre-existing dataset $D$. Our task evolves from subset
selection to the creation of an entirely new, ``domain-specific'' dataset from
scratch. 

\parh{Domain-Specific Context.}~In the context of this work, ``domain-specific''
refers to code related to particular applications or specialized programming
areas. For instance, relevant commercial products like PECAN~\cite{Pecan} specifically focus on synthesizing code for machine learning tasks (where machine learning is their focused domain). 
Nevertheless, establishing a universal definition over ``domain-specific'' code
appears challenging. Therefore, we adopt a pragmatic approach: we deem one
library representative enough to scope a domain. For instance, we consider NumPy
as representative of the scientific computing domain, Pandas for data
manipulation and analysis, and Matplotlib for data visualization. This approach
is intuitive, and it allows us to concretely define and work with
domain-specific datasets throughout our study. Unless otherwise specified,
subsequent references to domain-specific contexts in this paper adhere to this
definition.

\begin{figure}[!t]
    \centering
    \begin{subfigure}[t]{0.7\textwidth}
    \centering
    \includegraphics[width=\textwidth]{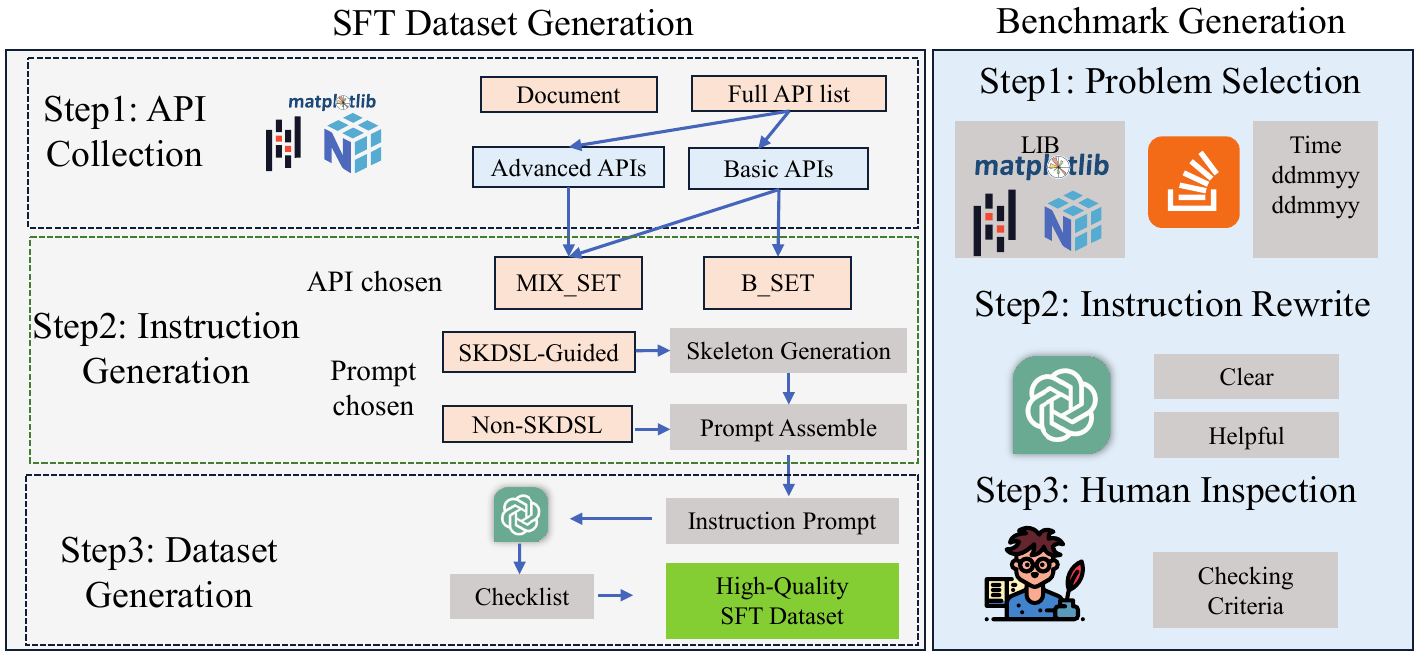}
    \caption{Overview of \toolg\ for SFT dataset generation and \bench\ for evaluation.}
    \label{fig:overview-TOOLG}
    \end{subfigure}
    \hspace{-0.00\textwidth}
    \begin{subfigure}[t]{0.25\textwidth}
    \centering
    \includegraphics[width=\textwidth]{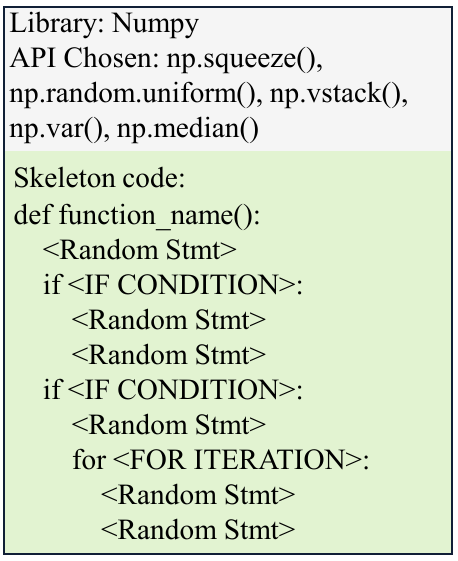}
    \caption{An example of key components in \toolg's step2.}
    \label{fig:skeleton-example}
    \end{subfigure}
    \vspace{-6pt}
    \caption{Overview of \toolg\ for SFT dataset generation and \bench\ for evaluation, with an example of key components in  \toolg's instruction generation process.}
    \vspace{-10pt}
    \label{fig:overview-TOOLG-big}
\end{figure}

\F~\ref{fig:overview-TOOLG} illustrates the three-step process employed by
\toolg\ for domain dataset synthesis: \ding{192} API collection: Given the
documentation of a library (usually can be obtained by crawling the library's official website), \toolg\ first extracts the complete API
list, then categorizes them into ``Advanced APIs'' and ``Basic APIs'' based on
their popularity and complexity. \ding{193} Instruction generation: \toolg\
further combines the APIs into $\text{B_SET}$ and $\text{MIX_SET}$ to specify
the target functionality of the generated code. It also incorporates skeleton
domain-specific language (SKDSL) to define the code structure,
resulting in the assembly of comprehensive instruction prompts.
\ding{194} Dataset generation: Finally, \toolg\ uses the instruction prompts to
generate data, validates them against a predefined checklist, and produces a
high-quality SFT dataset. These three steps are elaborated in
\S~\ref{subsubsec:api-collection}, \S~\ref{subsubsec:instruction-generation},
and \S~\ref{subsubsec:dataset-generation}, respectively.

In the following sections, we apply \toolg\ to three distinct domains: scientific computing, data manipulation and analysis, and data visualization. These domains are represented by NumPy, Pandas, and Matplotlib, respectively, chosen for their comprehensive API collections and meticulously maintained codebases. This
selection allows us to demonstrate the versatility and effectiveness of \toolg\
across a range of practical programming scenarios. Furthermore, it is important to note that the proposed technique is general and can be applied to other well-documented domains.

\parh{SFT Data Generation vs. Test Case Generation.}
The generation of high-quality SFT datasets shares certain commonalities with test case generation in software testing, despite their divergent objectives. While test case generation aims to produce scenarios that trigger software failures, SFT dataset generation seeks to create exemplars for improved model fine-tuning. In test case generation, complex cases that traverse deeper code branches are generally considered more valuable than trivial ones~\cite{fraser2011evosuite, pacheco2007randoop}. Similarly, in SFT dataset generation, the quality of instructions and code examples takes precedence over mere quantity. Zhou et al. \cite{zhou2024lima} demonstrate that utilizing 1,000 high-quality, manually crafted instructions by domain experts yields significantly greater performance improvements compared to larger but less curated datasets for SFT. However, a key distinction lies in the emphasis on diversity. For SFT datasets, achieving a diverse range of functional code completions is more crucial than solely pursuing complex corner cases.

\subsection{Generation Algorithm}
\label{subsec:toolg-methodology}

This section details \toolg's three-step process for generating high-quality SFT
datasets: API collection, instruction generation, and dataset generation. We
first introduce the key insights of \toolg's design, followed by giving the
details of each step.

\subsubsection{Key Insights}
\label{subsubsec:keyinnovation-toolg} 

\toolg\ incorporates two key insights. First, it frames the task as a
transformation process from high-level requirements to concrete implementations.
By providing the LLM with specific, comprehension-friendly functional and structural requirements for code
generation, \toolg\ reduces the need for extensive domain-specific knowledge,
enabling the model to focus on translating clear specifications into actual
code. Second, \toolg\ decomposes complex generation problems into simpler
sub-problems using API sets and SKDSL, lowering the capability threshold
required at each step. 

As a result of these insights, \toolg\ reduces the strict demand for a
``high-capability'' model, potentially enabling a
\textit{weak-to-strong}~\cite{burns2023weak} generation paradigm where less
capable models could be used for initial data generation, which is then used to
further improve the model performance. \toolg's novel approach enhances
flexibility and cost-effectiveness, establishing it as a versatile and scalable
framework for specific dataset synthesis, adaptable to varying model
capabilities and domain requirements.

\subsubsection{API Collection} 
\label{subsubsec:api-collection}

To use \toolg\ for dataset generation, we first extract APIs from a targeted
library specified by the user. As illustrated in \F~\ref{fig:overview-TOOLG}, we
first extract APIs from the library's official documentation, following the
method proposed by \cite{chen2021evaluating}. This process yields 4,089, 3,296,
and 3,683 APIs from Matplotlib, Pandas, and Numpy respectively, including their
parameters and functional descriptions.
Then, we rule out certain APIs based on the following criteria: (1) APIs
starting with ``__'' or ``c.'' are excluded, as they are typically internal or
low-level APIs related to the underlying C implementation; (2) For method
implementations in both base and derived classes, we only retain the base class
implementation as the API call; (3) any API invocation with a method
chain~\cite{nakamaru2020empirical} longer than three is excluded from further
consideration. 

Finally, we categorize the kept APIs into basic and advanced types. We select
basic APIs directly from the examples in the official tutorials of each library,
limiting their number to 50. These basic APIs are frequently used and easily
understood, such as \texttt{numpy.sum} and \texttt{pandas.read_csv}. The
remaining APIs are classified as advanced, which are typically less common and
often require specialized knowledge, as exemplified by \texttt{numpy.linalg.eig}
and \texttt{pandas.DataFrame.groupby}.

\parh{Proprietary Libraries.}~Notably, while this extraction and filtering process is straightforward for most public libraries with comprehensive documentation, we assume similar completeness for proprietary libraries considered for SFT dataset construction. It is worth mentioning that only high-quality internal codebases typically require SFT, and organizations with substantial codebases may customize their filtering rules to best suit their specific needs and codebase characteristics.

\subsubsection{Instruction Generation}
\label{subsubsec:instruction-generation}

As shown in \F~\ref{fig:overview-TOOLG}, the second step of \toolg\ involves
preparing instructions for querying the LLM to synthesize the SFT dataset. More
specifically, this step is divided into two separate parts: choosing the
appropriate APIs and selecting the prompt template. For the chosen APIs, we
consider two different strategies to create API sets of varying difficulty
levels:

\begin{equation}
\begin{split}
\text{B_SET} = \text{RandomSample}(\text{BasicAPIList}, N) \\
\text{MIX_SET} = \text{RandomSample}(\text{BasicAPIList} \cup \text{AdvancedAPIList}, N)
\end{split}
\end{equation}

\begin{table}[t]
    \caption{The prompt template with two slots (\textcolor[rgb]{0,0,0.9}{\{Library\}} and \textcolor[rgb]{0,0,0.9}{\{API chosen\}}). Text in \textcolor[rgb]{0.9,0,0}{red} will only show up when SKDSL is enabled.}
    \vspace{-10pt}
    \small
    \begin{tcolorbox}

    [System prompt]: You are a teacher who is good at \textcolor[rgb]{0,0,0.9}{\{Library\}}. You are exceptionally skilled at crafting high-quality programming problems and offering precise solutions.
    \\

    [User prompt]: Please take inspiration from the following list of application interfaces and their definitions to create a quality programming problem. Requirement: Use to all APIs in the list. Present your output in two distinct sections: [Problem Description] and [Solution]. 
    API list for inspiration: \textcolor[rgb]{0,0,0.9}{\{API chosen\}}
    \textcolor[rgb]{0.9,0,0}{You will be given a Python code skeleton, and you need to follow the structure to complete your solution. Example Python code skeleton: \{SKDSL code\}}
    \\
    Guidelines for each section: 1. [Problem Description]: This should be **completely self-contained**, providing all the contextual information one needs to understand and solve the problem. Assume common programming knowledge, but ensure that any specific context, variables, or code snippets pertinent to this problem are explicitly included. 
    2. [Solution]: Offer a comprehensive, **correct** solution that accurately addresses the [Problem Description] you provided.
    
\end{tcolorbox}
\label{tab:prompt-template}
\vspace{-10pt}
\end{table}
$\text{B_SET}$ focuses on commonly used APIs, ensuring high coverage of basic
functionalities. $\text{MIX_SET}$ introduces higher difficulty by incorporating
advanced APIs, simulating real-world scenarios where complex APIs are used
alongside basic ones. The selected APIs, along with their relevant information
extracted from documentation, are then incorporated into the prompt template
(see \T~\ref{tab:prompt-template}).

\parh{SKDSL for Code Skeleton Generation.}~While API sets of varying
difficulties control code content complexity, we design SKDSL, a domain-specific
language, to govern code structure by allowing rapid prototyping of Python's
high-level logic flow through specified Python keywords (e.g., \texttt{def},
\texttt{if}, \texttt{else}).
Given a randomly generated keyword list, we incrementally incorporate each
keyword into the code, randomly injecting valid statements (e.g., \texttt{a =
1}) between keywords. These injected statements serve to create a more complete
code skeleton, enabling subsequent validation by a syntax checker. While this
approach may occasionally produce invalid code skeletons, we maintain it for its
simplicity and efficiency. If an invalid skeleton is detected, we discard it and
generate a new one. This process of generating and validating is significantly
faster than querying the LLM for generation, often by a factor of thousands.
Consequently, the occasional generation of invalid skeletons has a negligible
impact on overall efficiency, while allowing us to produce a diverse range of
valid code skeletons.

After generating the code skeletons, SKDSL integrates with a grammar checker to
perform basic validation, catching syntactic errors early before full
implementation. This process enhances the quality of the generated skeletons.
Subsequently, we standardize these skeletons by replacing random statements with
the special token \texttt{<Random Stmt>} and conditions in control flow keywords
(e.g., \texttt{if}, \texttt{while}) with \texttt{<Corresponding Keyword +
Condition>}. These refined and standardized code skeletons then serve as the
example code part in the prompt template (see \T~\ref{tab:prompt-template}),
which is ultimately assembled into the final prompt for LLM querying. To
illustrate this process, we present an example in \F~\ref{fig:skeleton-example},
where a skeleton code incorporating two \texttt{if} statements and one
\texttt{for} loop is generated using five chosen NumPy APIs, demonstrating how
these elements are assembled to create the final prompt.

\subsubsection{Dataset Generation}
\label{subsubsec:dataset-generation}

The final stage of \toolg\ uses the generated instruction prompts to query the LLM, producing library-specific instruction and response code pairs. Each pair undergoes format and length checks, discarding those that fail to meet specified criteria. These criteria include the presence of a code snippet and content length requirements, where pairs with fewer than 32 tokens or more than 4,096 tokens are excluded.
Subsequently, we perform content validation on the remaining pairs. As mentioned in~\S~\ref{subsubsec:instruction-generation}, the inserted random valid statements enable grammatical correctness checks. Based on this, we define a threshold T (0 < T < 1) to determine code acceptance. If the number of detected APIs in the generated code exceeds $N \times T$, where N is the required number of APIs, the pair passes content validation. Only the pairs that successfully pass all the aforementioned checks are included in our SFT dataset.

\parh{Automated Pipeline.}~Notably, \toolg\ is designed to establish an efficient and automated SFT dataset generation process. The entire procedure does not necessitate any manual intervention, from API collection through instruction generation to dataset creation. It guarantees the scalability and reproducibility of the dataset generation pipeline, facilitating the creation of extensive, high-quality SFT datasets for a wide range of domain-specific libraries. By minimizing human involvement, we not only increase efficiency but also reduce the potential for human-induced biases or errors, ensuring consistent quality across large-scale dataset synthesis efforts.

\subsection{\bench: A Benchmark for Evaluating Code Generation in Specific Domains}
\label{subsec:toolg-benchmark}

To evaluate the effectiveness of \toolg\ and address the lack of existing
benchmarks in this area, we developed \bench, a novel benchmark designed to
assess model performance on specific domains. As outlined in
\F~\ref{fig:overview-TOOLG}, the creation of \bench\ follows a three-step
approach: \ding{192} problem selection through a highly automated collection
process, \ding{193} instruction rewriting using an automated rewriting process,
and \ding{194} human inspection with rigorous selection criteria. 
This comprehensive process ensures the quality and relevance of the included questions, facilitating further research and assessment of LCMs in specific domains.
In the following subsections, we detail each step of this process, beginning with our approach to problem selection.

\subsubsection{Problem Selection}
\label{subsubsec:problem-selection}
We use the Stack Exchange platform~\cite{stackexchange} to identify popular library-specific questions, searching with the target library name as the keyword. We prioritize questions with accepted answers or, in their absence, those with the highest-voted responses. To ensure quality and relevance, we apply a filtering criterion based on the average monthly vote count, considering only questions with at least 5 votes per month. We also verify the presence of the target library in the answer code snippets. This process yields 439, 642, and 391 Python-related question-answer pairs for Numpy, Matplotlib, and Pandas, respectively, encompassing a diverse range of domain-specific tasks.

\parh{Data Leakage.}~Notably, to mitigate the risk of data leakage and maintain the integrity of the evaluation, we exclusively consider questions posted between August 2023 and April 2024, ensuring that the selected data is temporally distinct from the training data of the LCM model used for comparison in our experiments. This temporal separation guarantees that the benchmark accurately assesses the model's ability to generalize to unseen domain-specific tasks, providing a fair and unbiased evaluation of its performance.

\subsubsection{Instruction Rewriting} 
\label{subsubsec:instruction-rewriting}
To enhance the quality of the selected question-answer pairs, we employ a rewriting process similar to previous works~\cite{zhou2024lima,liu2024coachlm}. First, we extract code snippets from raw answers, removing unrelated information, indentation symbols, and HTML tags (e.g., \texttt{<p>}). Then, we utilize GPT-4 to generate refined instruction pairs based on the questions and answers. This process involves improving clarity, eliminating unnecessary details, and ensuring the instructions are easily comprehensible. GPT-4 assists in creating clear, concise guidance accompanied by relevant code snippets. The result is a set of high-quality instruction pairs that effectively capture the essence of the original question-answer pairs while significantly enhancing their clarity and usefulness for evaluating LCMs in specific libraries.

\begin{table}[!htbp]
    \centering
    \caption{Human evaluation criteria for the quality of instruction pairs.}
    \vspace{-10pt}
    \label{tab:human-eval-criteria}
    \resizebox{1.0\linewidth}{!}{
    \begin{tabular}{lp{8cm}p{8cm}}
    \hline
    \multicolumn{3}{c}{\textbf{Criteria for INSTRUCTION}} \\
    \hline
    \textbf{Dimension} & \textbf{Description} & \textbf{Main Checklist} \\
    \hline
    Contextualization & The instruction includes context on code functionality, inputs/outputs, and constraints. & Check for clear specs of code purpose, interface, requirements and key details. \\
    \hline
    Feasibility & The instruction is clear, specific, feasible, and easily understandable. & Check for ambiguity, errors or unreasonable requests beyond AI's capability. \\
    \hline
    Readability & The instruction follows conventions for describing software requirements. & Check for terminology, completeness, consistency, and organization. \\
    \hline
    Relevance   & The instruction is relevant to the target library. & Check that the task aligns with the intended use and functionality of the target library's APIs.  \\
    \hline
    \multicolumn{3}{c}{\textbf{Criteria for RESPONSE}} \\
    \hline
    \textbf{Dimension} & \textbf{Description} & \textbf{Main Checklist} \\
    \hline
    Correctness & Responses should be syntactically correct, functional, and follow the library's best practices. & Check for syntax errors, logical bugs, and deviations from the library's common usage or style guidelines. \\
    \hline
    Readability & Responses should be clean, well-structured, and use meaningful variables, avoiding excessive randomness. & Check for unclear code, inconsistent formatting, and poorly named variables or functions. \\
    \hline
    Usefulness & Responses should directly address the user's request and provide helpful solutions. & Check that the code directly solves the requested task and provides a helpful solution. \\
    \hline
    Safety & Responses should be harmless and not about non-ethical topics. & Check for the promotion of illegal activities or inappropriate content.  \\
    \hline
    \end{tabular}
    }
    \vspace{-10pt}

\end{table}

\subsubsection{Human Inspection}
\label{subsubsec:human-inspection}
Finally, in the third step, human inspection, we carefully review the rewritten instruction pairs to ensure their quality and relevance to the target library. To maintain consistency and objectivity during the human inspection process, we adhere to a set of predefined criteria, as detailed in \T~\ref{tab:human-eval-criteria}. These criteria encompass various aspects, including contextualization, feasibility, readability, relevance, correctness, usefulness, and safety. Our inspection process follows a similar approach to that of \cite{liu2024coachlm}, with adaptations made to accommodate the specific requirements of code generation. By applying these criteria rigorously, we ensure that the final benchmark dataset meets high standards of quality and can effectively evaluate the performance of LCMs in specific libraries. Instruction pairs that fail to satisfy any of the specified criteria are excluded from the final benchmark. This meticulous inspection process guarantees that only the most suitable and relevant instruction pairs are included. As a result of this comprehensive evaluation, we obtain 115 high-quality instruction pairs for each library. This curated benchmark dataset covers a wide range of domain-specific tasks, providing a robust foundation for assessing the capabilities of LCMs in generating accurate and effective code snippets for domain-specific programming challenges.

\subsection{Experimental Setup for Specific Scenario}
\label{subsec:toolg-setup}

To evaluate the effectiveness of \toolg\ on \bench, we employ the same models described in \S~\ref{subsec:tools-experiment}. However, to better align with the specific requirements of \bench, we have adjusted our evaluation metrics and some hyperparameters accordingly.

\parh{Metrics.}~We employ four distinct metrics to comprehensively assess the performance of \toolg:

\begin{itemize}
    \item \textbf{CodeBLEU}: We use CodeBLEU~\cite{ren2020codebleu}, a metric that evaluates the quality of generated code snippets by comparing them to the ground truth code snippets. CodeBLEU takes into account token-level, structural-level, and semantic-level information. It consists of four components: n-gram matching score $BLEU$, weighted n-gram matching score $weighted\_BLEU$, syntactic AST matching score $AST\_Score$, and semantic data flow matching score $DF\_Score$. The overall CodeBLEU score is calculated as:
    
    \begin{equation}
    \label{equ:codebleu}
    \begin{aligned}
        CodeBLEU &= \alpha * BLEU + \beta * weighted\_BLEU \\
                 &+ \gamma * AST\_Score + \delta * DF\_Score
    \end{aligned}
    \end{equation}
    
    \noindent where $\alpha, \beta, \gamma, \delta$ are the weights for each component, all set to 0.25 as suggested in~\cite{wang2021codet5, lu2021codexglue}. A higher CodeBLEU score indicates better quality of the generated code snippets.

    \item \textbf{Cyclomatic Complexity}: We use cyclomatic complexity~\cite{mc1977cycle} to measure the complexity of the code generated by \toolg. This metric corresponds to the number of linearly independent paths through the code and serves as an indicator of the code's complexity, which quantifies the number of decisions within a block of code. We apply Radon~\cite{radon} to analyze the AST tree of a Python program and compute its cyclomatic complexity.

    \item \textbf{Silhouette Coefficient}: This metric measures the similarity of a sample point to its own cluster compared to other clusters. It is calculated as $(b(i) - a(i)) / \max(a(i), b(i))$, where $a(i)$ is the average distance between sample point $i$ and all other points in the same cluster, and $b(i)$ is the minimum average distance between $i$ and points in any other cluster. The value ranges from -1 to 1, with higher values indicating better clustering results. The overall Silhouette Coefficient is the mean of the coefficients for all sample points.

    \item \textbf{Calinski-Harabasz Index}: Similarly, Calinski-Harabasz Index
    (CH-Index) is used to assess the compactness and separation of clusters. It
    is calculated as $(SSB / (k - 1)) / (SSW / (n - k))$, where $SSB$ is the
    between-cluster sum of squares, measuring the variance between cluster
    centers, and $SSW$ is the within-cluster sum of squares, measuring the
    variance of sample points within each cluster. $k$ is the number of
    clusters, and $n$ is the total number of sample points. A higher
    Calinski-Harabasz Index indicates better clustering quality, with more
    compact clusters and greater separation between them.
\end{itemize}

\parh{Hyperparameters.}~We maintain consistency with most of the hyperparameter settings outlined in \S\ref{subsec:tools-experiment}. However, to enhance training stability and ensure reliable results, we adjust the training batch size from 64 to 32 for datasets containing fewer than 2,000 examples.

\subsection{Results for \toolg}
\label{subsec:evalution-toolg}

In this section, we present a comprehensive evaluation of \toolg. We begin by
examining its performance on \bench\ and analyzing the impact of dataset size.
Subsequently, we delve into an analysis of internal logits for clustering
evaluation. We then investigate the relationship between pass rate and cost, as
well as the influence of hyperparameters. Finally, we employ both LLM-based and
human evaluations to assess the quality of \toolg's output.

\subsubsection{Main Results}
\label{subsubsec:main-results-bench}

\begin{table*}[!t]
    \centering
    \caption{Performance comparison of 7B and 13B models on the \bench\ dataset.
    Scores are reported using the CodeBLEU metric. Bold values indicate the highest
    scores for each model size and library combination.}
    \vspace{-5pt}
    \label{tab:TOOLG-results}
    \resizebox{0.7\linewidth}{!}{
    \begin{tabular}{llccccc}
    \hline
    Model Size & Dataset & DSL-Guided & \# Examples & Numpy & Pandas & Matplotlib \\ \hline
    13B & - & - & -                    & 0.1917 & 0.2014 & 0.2177 \\
    13B & CODEE & -  & 76k             & 0.3218 & 0.3183 & 0.3436 \\
    13B & MIX                & No & 2k & 0.3380 & 0.3151 & 0.3679 \\
    13B & BASIC              & No & 2k & 0.3493 & 0.3319 & 0.3699 \\
    13B & COMB               & No & 4k & 0.3563 & 0.3470 & 0.3722 \\ \hline
    13B & BASIC              & Yes &2k & 0.3546 & 0.3359 & 0.3472 \\ 
    13B & MIX                & Yes &2k & 0.3535 & 0.3314 & 0.3519 \\ 
    13B & COMB               & Yes &4k & 0.3681 & 0.3452 & 0.3513 \\ \hline
    13B & COMB-BOTH          & Yes &4k & \textbf{0.3815} & \textbf{0.3552} & \textbf{0.3973} \\ \hline
    7B & - & - & -                     & 0.1467 & 0.1327 & 0.1442 \\
    7B & CODEE & - & 76k               & 0.3112 & 0.2994 & 0.3312 \\ 
    7B & MIX       & No & 2k           & 0.3169 & 0.2802 & 0.3522 \\
    7B & BASIC       & No & 2k         & 0.3205 & 0.2811 & 0.3668 \\
    7B & COMB           & No & 4k      & 0.3324 & 0.3266 & 0.3714 \\ \hline  
    7B & MIX          & Yes  &  2k     & 0.3402 & 0.3064 & 0.3443 \\ 
    7B & BASIC          & Yes  &2k     & 0.3392 & 0.3033 & 0.3324 \\ 
    7B & COMB           & Yes  &4k     & 0.3425 & 0.3474 & 0.3336 \\ \hline
    7B & COMB-BOTH      & Yes  &4k     & \textbf{0.3452} & \textbf{0.3544} & \textbf{0.3795} \\ \hline
    \end{tabular}
    }
    \vspace{-15pt}

\end{table*}

\T~\ref{tab:TOOLG-results} presents the evaluation results on \bench\ using
CodeBLEU as the metric for CodeLlama models of 7B and 13B sizes. We compare the
performance of base models without fine-tuning, models fine-tuned on the full
CODEE dataset (76K examples), and models fine-tuned using our generated
datasets. ``DSL-Guided'' indicates whether SKDSL was used in dataset generation.
The ``Dataset'' column specifies the prompt selection method: BASIC uses only
the $\text{B_SET}$, MIX uses only the $\text{MIX_SET}$, COMB combines both BASIC
and MIX, and COMB-BOTH randomly selects 2K examples from each of the two COMB
datasets, resulting in a total of 4K examples.  The ``\# Examples'' column shows
the detailed number of examples in each SFT dataset.

The results in \T~\ref{tab:TOOLG-results} demonstrate that fine-tuning the
models using our generated datasets consistently improves their performance on
all three libraries compared to the base versions without fine-tuning. For
instance, fine-tuning the 13B and 7B models with COMB-BOTH (4K examples)
achieves CodeBLEU scores of 0.3973 and 0.3795 on Matplotlib, representing
relative improvements of 82.4\% and 163.2\% over their respective base models.
Moreover, using COMB-BOTH achieves better results than fine-tuning with the full
76K examples from the CODEE dataset. For example, fine-tuning the 13B model with
COMB-BOTH outperforms fine-tuning with CODEE by 9.3\%, 11.6\%, and 15.6\% on
Numpy, Pandas, and Matplotlib, respectively.

\parh{Impact of API Selection Strategies.} Both API selection strategies
contribute to the improvement in model performance. The models fine-tuned with
COMB consistently exhibit higher scores compared to each component across all
three repositories. On average, fine-tuning the 13B model with COMB improves the
CodeBLEU score by 3.5\% and 5.1\% compared to fine-tuning with MIX and BASIC,
respectively. While some individual improvements may appear modest in absolute
terms, the consistent upward trend across various settings underscores that the
coverage of API combinations is crucial to enhance the models' ability to
generate accurate and relevant code.

\parh{Impact of SKDSL.} SKDSL plays a vital role in generating high-quality SFT datasets. Using SKDSL improves the CodeBLEU score by 7.3\% and 5.2\% on Numpy and Pandas, respectively, compared to the Non-SKDSL approach. However, for Matplotlib, the Non-SKDSL SFT dataset yields better performance. Analysis of code length and cyclomatic complexity reveals that SKDSL-generated code is longer (37.51\%, 43.57\%, and 27.01\% for Numpy, Pandas, and Matplotlib) and more complex (71.68\% higher cyclomatic complexity) than Non-SKDSL code. 

We hypothesize that this discrepancy in performance across libraries is due to their inherent characteristics. The increased complexity benefits libraries like Numpy and Pandas, which require more logical reasoning. In contrast, Matplotlib often involves simpler workflows where complex code structures are less common. It's worth noting that we did not dynamically adjust SKDSL specifications to generate simpler structures in our experiments, which might have influenced the results for Matplotlib. Despite these variations, COMB-BOTH achieves the best performance across all three repositories, outperforming the two COMB groups by an average of 5.7\% and 3.2\% on the 7B and 13B models, respectively. These findings demonstrate the effectiveness of \toolg\ in generating SFT datasets, emphasizing the importance of considering both API coverage and code structure in the dataset generation process.

\parh{Impact of Dataset Size.}~Audiences may question our choice of 2k as the
minimum dataset size in the above experiments. The primary rationale behind this
decision stems from previous research~\cite{zhou2024lima}, which suggests that
2k high-quality examples are sufficient to produce an effective SFT dataset and
yield an excellent model with a stable training process. Conversely, using too
few examples (e.g., 10) can lead to training difficulties, such as convergence
issues.

\begin{table}[!t]
    \caption{Impact of dataset size on model performance and API coverage on Numpy.}
    \vspace{-5pt}
    \centering
    \resizebox{0.55\linewidth}{!}{
    \begin{tabular}{lccccc}
    \hline
    Dataset Size & 1k & 2k & 3k & 4k \\ \hline
    CodeBLEU & 0.3142 & 0.3169 & 0.2969 & 0.3006 \\
    Basic API coverage & 1.00 & 1.00 & 1.00 & 1.00 \\
    Advanced API coverage & 0.9106 & 0.9525 & 0.9777 & 0.9804\\
    \hline
    \end{tabular}
    \vspace{-10pt}
    \label{tab:dataset-size}
    }
\end{table}

To further validate this choice, we conduct experiments on Numpy using four
different dataset sizes: 1k, 2k, 3k, and 4k. The results are presented in
\T~\ref{tab:dataset-size}.
Our findings reveal that a 1k dataset achieves 100\% coverage for basic APIs,
while a 2k dataset surpasses 95\% coverage for advanced APIs. The CodeBLEU
scores for the 1k, 2k, 3k, and 4k datasets are 0.3142, 0.3169, 0.2969, and
0.3006, respectively. Interestingly, we observe that the CodeBLEU score peaks at
2k and slightly decreases for larger dataset sizes. The coverage of advanced
APIs increases from 0.9106 for the 1k dataset to 0.9804 for the 4k dataset, but
the rate of improvement slows down as the dataset size grows beyond 2k.
Balancing computational costs with the saturation of API coverage, we conclude that a 2k dataset size offers an optimal trade-off between model performance and efficiency.

\subsubsection{Logits Analysis and Clustering Evaluation}
\label{subsubsec:logits-analysis}

Beyond evaluating the performance improvement of our supervised fine-tuned
models on downstream code generation tasks, we conduct a deeper analysis to
investigate how \toolg-generated SFT datasets impact the models' ability to
distinguish between different third-party libraries. We treat questions from
\bench\ pertaining to distinct third-party libraries as separate categories. Our
analysis process involves several steps: \ding{192} encoding the input
instructions using the tokenizer to obtain input IDs and attention mask tensors;
\ding{193} performing a forward pass through the specific model with the encoded
inputs to obtain the logits output; \ding{194} computing the sum of logits
weighted by the attention mask from the last layer; and \ding{195} applying
t-Distributed Stochastic Neighbor Embedding (t-SNE)~\cite{van2008visualizing} to
reduce the dimensionality of the logits sum data to two dimensions for
visualization.

We evaluate clustering quality using Silhouette score and CH Index score (See
in~\S~\ref{subsec:toolg-setup}), analyzing the base model and two fine-tuned
variants for each model size. \T~\ref{tab:average-scores} presents the average
scores for each model configuration.

\begin{table}[htbp]
    \vspace{-5pt}
    \caption{Average Silhouette and CH Index scores for each model.}
    \vspace{-10pt}
    \centering
    \resizebox{0.75\linewidth}{!}{

    \begin{tabular}{l|l|c|c|c}
    \hline
    Base Model & Dataset & \# Examples & Avg. Silhouette & Avg. CH Index \\
    \hline
    CodeLlama-7b & - & - & 0.079 & 6.282 \\
    CodeLlama-7b & CODEE & 76k & 0.085 & 6.249 \\
    CodeLlama-7b & COMB-BOTH & 4k & 0.076 & 6.906 \\
    CodeLlama-13b & - & - &  0.287 & 80.413 \\
    CodeLlama-13b & CODEE& 76k & 0.120 & 14.107 \\
    CodeLlama-13b & COMB-BOTH & 4k & 0.315 & 99.745 \\
    \hline
    \end{tabular}
    }
    \label{tab:average-scores}
\end{table}

\begin{figure}[!t]
    \centering
    \begin{subfigure}{0.32\textwidth}
    \centering
    \includegraphics[width=\textwidth]{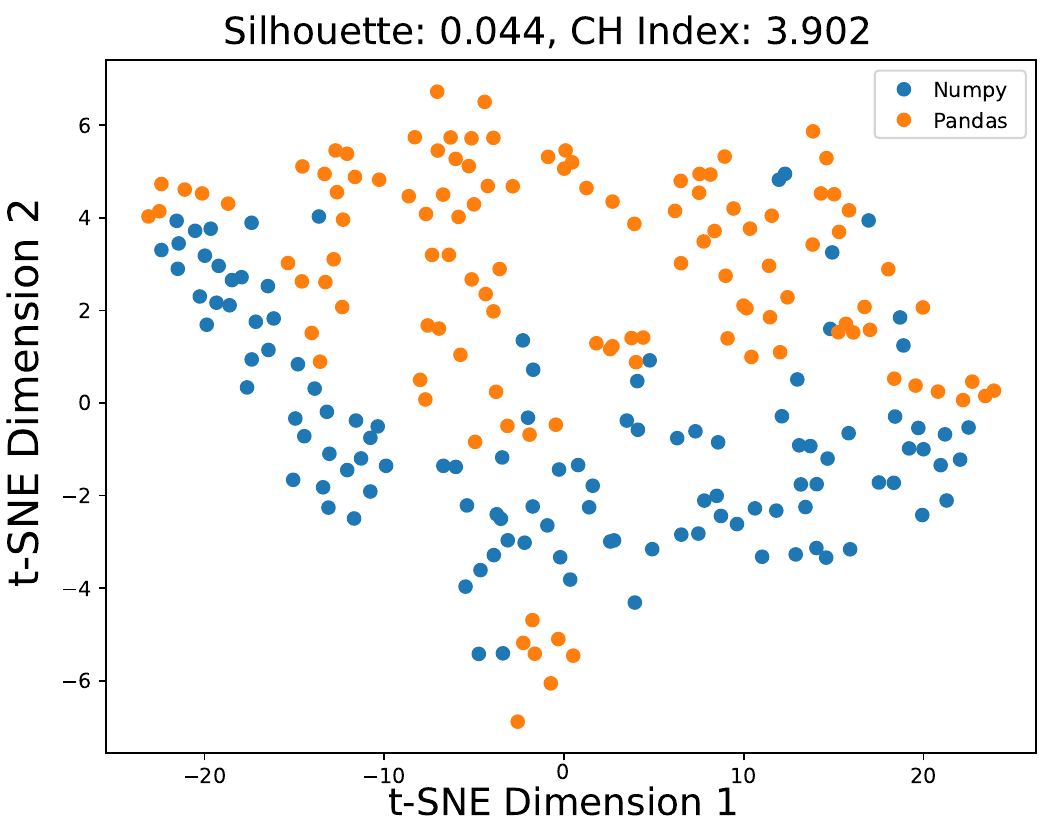}
    \caption{7B COMB-BOTH model}
    \label{fig:anomalous-clustering}
    \end{subfigure}
    \hspace{-0.01\textwidth}
    \begin{subfigure}{0.32\textwidth}
    \centering
    \includegraphics[width=\textwidth]{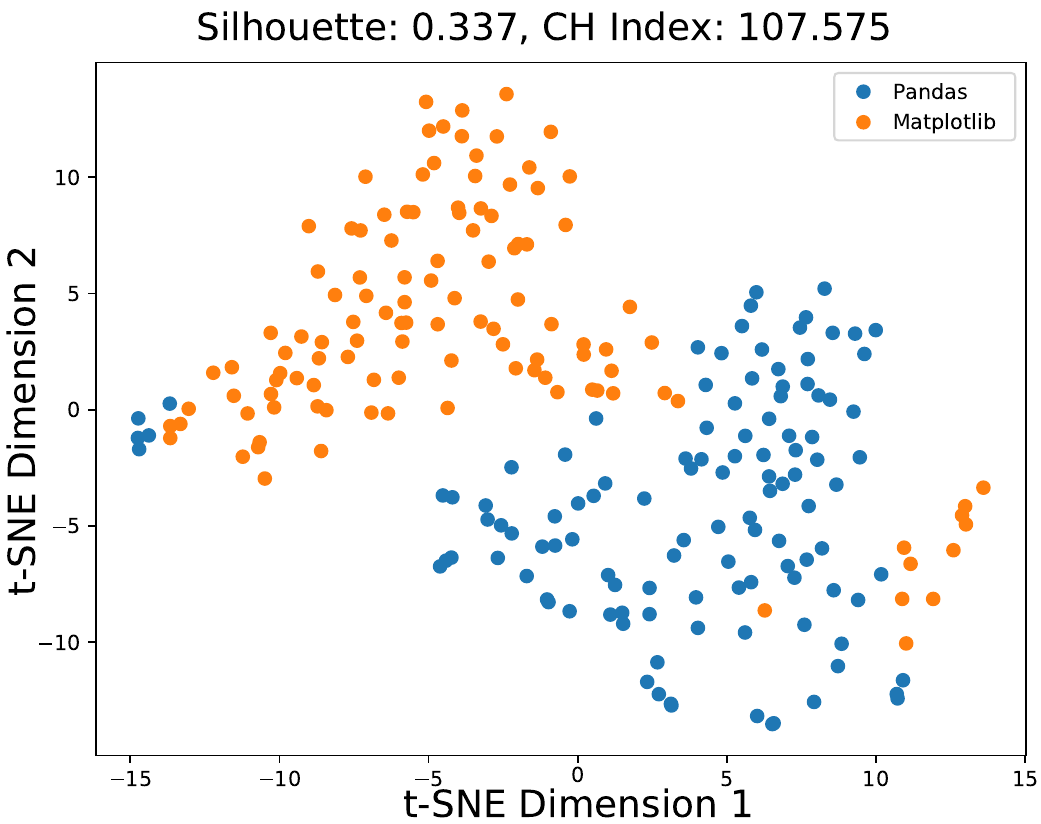}
    \caption{Base 13B model}
    \label{fig:base-model}
    \end{subfigure}
    \hspace{-0.01\textwidth}
    \begin{subfigure}{0.32\textwidth}
    \centering
    \includegraphics[width=\textwidth]{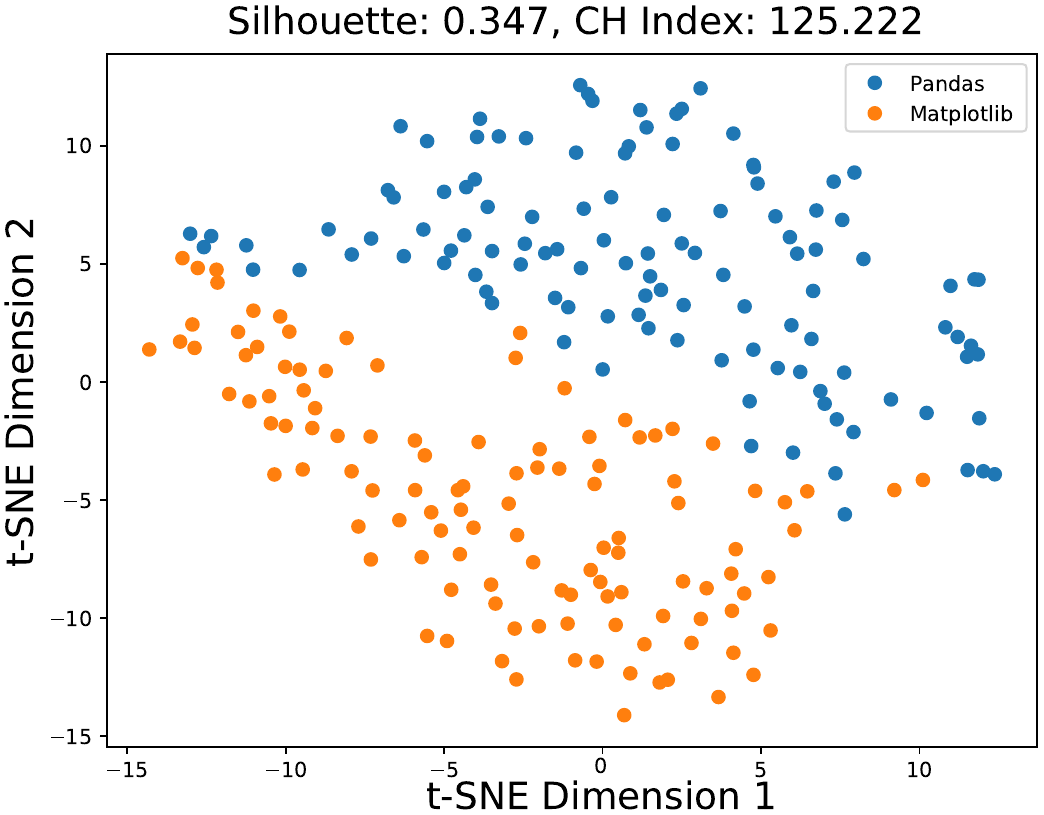}
    \caption{13B COMB-BOTH model}
    \label{fig:fine-tuned-model}
    \end{subfigure}
    \vspace{-6pt}
    \caption{(a) Clustering behavior in the 7B model. (b) and (c) Comparison of the base 13B model and model fine-tuned with \toolg\ in distinguishing between Matplotlib and Pandas.}
    \vspace{-10pt}
    \label{fig:combined-figure}
\end{figure}

Our experimental results reveal two key insights. First, the \toolg\ fine-tuned model consistently outperforms both the base model and the full dataset fine-tuned model across various configurations. It achieves average CH index improvements of 16.98\% and 408.76\%, respectively. Interestingly, for the 7B model, we observe a higher CH index despite a slightly lower Silhouette score. This anomaly stems from a small cluster of Pandas data points incorrectly grouped with Numpy, as illustrated in \F~\ref{fig:anomalous-clustering}.

Second, model size significantly impacts clustering performance. The 13B base model exhibits stronger discriminative power compared to the 7B model. Moreover, the \toolg\ fine-tuned model demonstrates an improved ability to distinguish between third-party libraries. \F~\ref{fig:base-model} and \F~\ref{fig:fine-tuned-model} illustrate this enhancement, showing the base and fine-tuned models' performance in distinguishing Matplotlib from Pandas. The fine-tuned model's improved discriminative performance aligns with the observed increases in Silhouette and CH index values.

\subsubsection{Pass Rate and Cost}
\label{subsec:passrate-eval}

Our initial evaluation employs GPT-3.5 as the backbone LLM for generating SFT
datasets, aligning with our baseline SFT
datasets~\cite{weimagicoder,CodeExercise} for fair comparison. However,
considering that vendors who construct these datasets and fine-tune models on
them prefer to avoid relying on closed-source models to prevent potential
copyright disputes, it is crucial to assess the ability of open-source models to
generate high-quality datasets. Furthermore, as outlined
in~\S~\ref{subsec:toolg-methodology}, each generated code snippet undergoes a
series of compliance checks, introducing a trade-off between generation cost and
pass rate.

\begin{figure}[!t]
    \centering
    \begin{subfigure}[t]{0.58\textwidth}
    \centering
    \includegraphics[width=\textwidth]{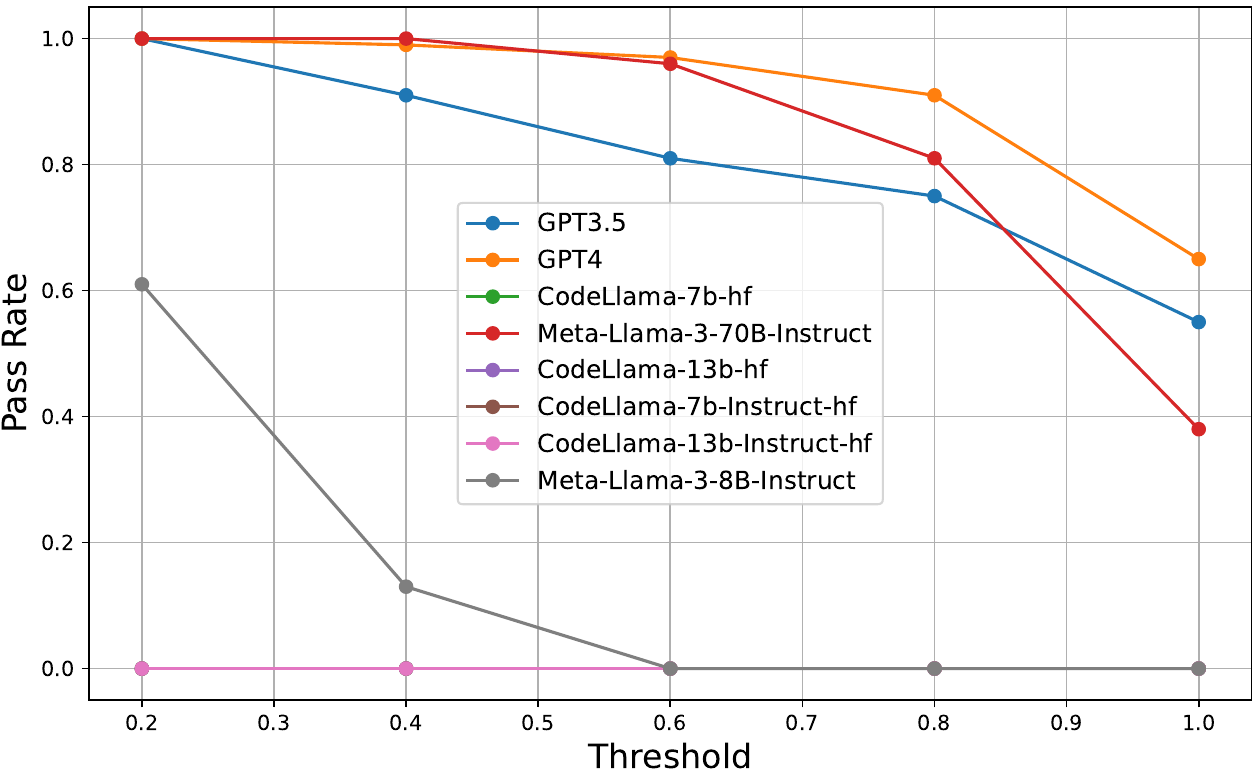}
    \caption{Comparative analysis of pass rates across various LLMs at different thresholds.}
    \label{fig:passrate}
    \end{subfigure}
    \hspace{-0.00\textwidth}
    \begin{subfigure}[t]{0.40\textwidth}
    \centering
    \includegraphics[width=\textwidth]{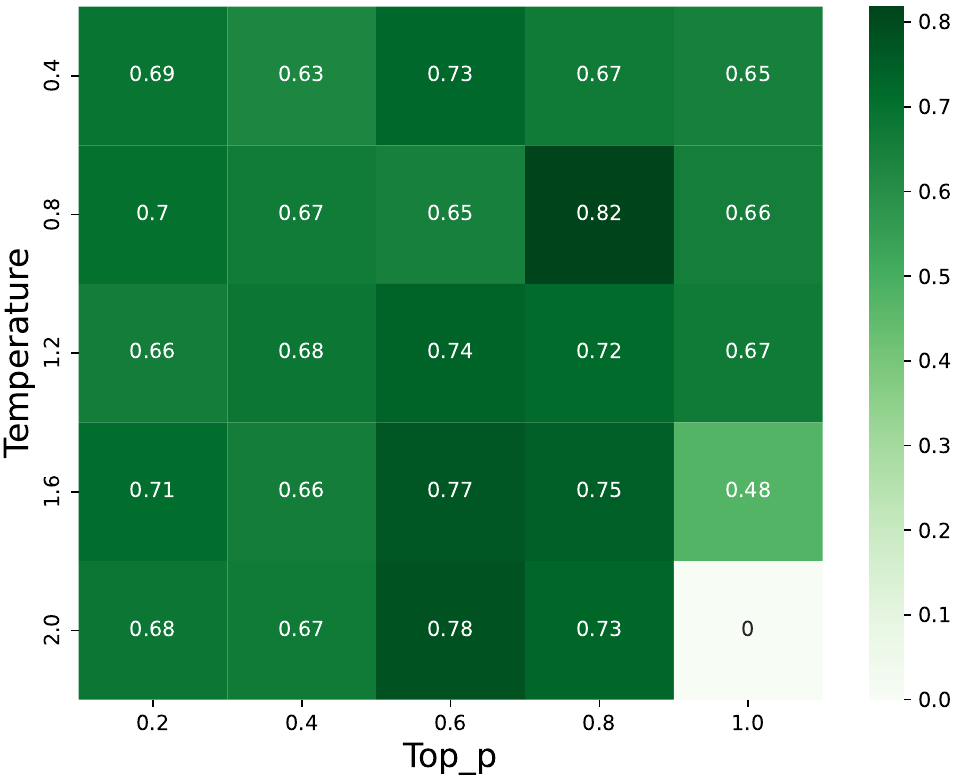}
    \caption{Heatmap visualization of pass rates under diverse hyperparameter configurations.}
    \label{fig:passrate-heatmap}
    \end{subfigure}
    \vspace{-6pt}
    \caption{Comprehensive evaluation of LLM performance: Pass rate analysis and hyperparameter impact assessment.}
    \vspace{-15pt}
    \label{fig:combined-passrate}
\end{figure}

To provide a comprehensive analysis of the trade-off between generation cost and
pass rate, we evaluate the pass rate of the SFT dataset generated by 8 different
models, including 6 open-source and 2 closed-source models of varying sizes. For
each model's generated responses, we set the threshold T, as described in
\S~\ref{subsec:toolg-methodology}, to (0.2, 0.4, 0.6, 0.8, 1.0) to calculate the
pass rates. The results, presented in \F~\ref{fig:passrate}, reveal a general
trend of decreasing pass rates as the threshold increases, indicating that more
stringent criteria lead to lower acceptance of generated data. However, the
performance varies among different models. In terms of pass rates alone, GPT-4
consistently outperforms all other models across all thresholds. Llama-3-70B,
the largest open-source model in our experiment, and GPT-3.5 alternate in
leading performance at different thresholds, highlighting their competitive
capabilities. Conversely, some smaller models (e.g., Llama-3-8B) or models that
have undergone code continuation pretraining (e.g., CodeLlama-13b) exhibit lower
pass rates across all thresholds.

Considering the price differences among the top three models in terms of pass
rates, GPT-3.5 emerges as a cost-effective option. The cost of generating one
million tokens using GPT-3.5 is 0.5 (input) / 1.5 (output), which is only 58\%
of the cost of Meta-Llama-3-70B and 5\% of the cost of GPT-4. Therefore, GPT-3.5
strikes a good balance between the generation cost and the pass rate, making it
a suitable choice for our evaluation and subsequent SFT. Our experiments show
that generating a 4k SFT dataset using GPT-3.5 costs approximately 3 USD,
whereas a human-written dataset of comparable scale and quality would require
contributions from over 334 volunteers~\cite{kopf2024openassistant}.

\subsubsection{Hyperparameter Impact}
\label{subsubsec:hyperparameter-impact}

To further investigate the impact of hyperparameters on the performance of \toolg, we conducted experiments with different combinations of temperature and top\_p values on GPT-3.5 with threshold T set to 0.6. 
The temperature ranged from 0.4 to 2.0, and top\_p ranged from 0.2 to 1.0. \F~\ref{fig:passrate-heatmap} shows the pass rate heatmap for various hyperparameter settings, where the pass rate value indicates the percentage of generated code that satisfies the specified API usage requirements.

The heatmap reveals that the pass rates range from 0.76 to 0.82 for most hyperparameter combinations, with the best performance achieved when both temperature and top\_p are set to 0.8, resulting in a pass rate of 0.82. This suggests that the default hyperparameter configuration (temperature=0.8, top_p=1.0) already yields good results, making it a suitable choice for generating high-quality data that adhere to the specified API usage requirements.
However, extreme hyperparameter configurations, such as setting both temperature and top\_p to very high values (e.g., temperature=2.0, top\_p=1.0), can lead to the generation of incoherent text with random characters. This is because a higher temperature value would increase the randomness of sampling, while a top\_p value of 1.0 considers all possible tokens during sampling. The combination of high temperature and top\_p significantly increases the uncertainty of the output, resulting in the generation of incoherent text containing a mix of seemingly random English words, programming terms, numbers, and non-standard Unicode symbols. These extreme settings can negatively impact the quality and coherence of the generated data, rendering them unusable for further fine-tuning.

\subsubsection{Pairwise Comparison of Model Performance}
\label{subsubsec:pairwise-comparison}

To further evaluate \toolg's effectiveness, we conduct pairwise comparisons between responses generated by different models using both LLM-based and human evaluations. We focus on 7B and 13B model sizes. For each size, we compare the COMB-BOTH version (our primary model) against four variants: the CodeLlama-provided instructed version, and versions fine-tuned on the full CODEE, BASIC, and MIX datasets respectively. All fine-tuned models use the settings described in \S~\ref{subsubsec:main-results-bench}.

\parh{LLM-based Evaluation.}
As advanced LLMs demonstrate superior performance in providing valuable evaluations, we follow previous work~\cite{zheng2024judging} and conduct an evaluation using GPT-4 as the judge for our pairwise comparison. For each question, we compare the responses generated by different models and ask GPT-4 to select the better one. The results are then aggregated to calculate the win rate of each comparison, as shown in the column ``Win Rate by GPT-4'' in \T~\ref{tab:gpt4evaluation}.

The results reveal two key findings. First, the model trained on the COMB-BOTH dataset consistently outperforms the instructed version and the model fine-tuned on the entire CODEE dataset, both of which utilize a significantly larger amount of data for fine-tuning. When using CodeLlama7b as the base model, COMB-BOTH achieves a remarkable win rate of 85.22\% against Instruct-hf and maintains win rates above 64\% across all comparisons. This observation highlights the effectiveness of \toolg\ in generating a small quantity of high-quality, domain-specific data to facilitate further SFT and improve performance in domain-specific code generation tasks.
Second, in line with the findings in \S~\ref{subsubsec:main-results-bench}, models fine-tuned with COMB-BOTH consistently exhibit a win rate of more than 50\% compared to each component (BASIC and MIX), demonstrating the importance of combining different API sets in generating high-quality SFT datasets.

\parh{Human Evaluation.}
To validate our LLM-based evaluation, we conduct a complementary human evaluation. We randomly select 25 samples for each model pair comparison and create an online questionnaire. We invite five experts, including two industrial developers and three academic researchers with expertise in Python, as participants. We provide two generated code snippets for the same question without specifying their origins and ask the participants to evaluate the quality of the code snippets based on the following criteria: (1) correctness, (2) relevance to the question, and (3) readability. The participants are then required to choose the better response.

The human evaluation results, presented in \T~\ref{tab:gpt4evaluation}, align with the LLM-based evaluation findings. The primary model achieves an average win rate of 85.30\%, significantly outperforming other models and reaffirming our approach's superiority in generating high-quality SFT datasets. To address potential inter-rater variability~\cite{peng1997validity}, we calculate the Fleiss' Kappa score~\cite{fleiss1971measuring} for the questionnaire. The resulting score of 0.63 indicates substantial agreement among participants, bolstering the reliability of our human evaluation results.

\begin{table}[!t]
    \caption{Win rates of model fine-tuned on COMB-BOTH compared to other models,
    evaluated by GPT-4 and human judges.}
    \vspace{-6pt}
    \centering
    \resizebox{0.75\linewidth}{!}{
    \begin{tabular}{c|c|c|c|c}
    \hline
    Base Model & Compared Model & \# Examples & Win Rate by GPT-4 & Win Rate by Human \\
    \hline
    \multirow{4}{*}{CodeLlama-7b} & Instruct-hf & - & 85.22\% & 94.67\%\\
    & CODEE&76k & 67.83\% & 74.52\% \\
    & BASIC&2k & 72.81\% &  88.15\%\\
    & MIX&2k & 69.30\% & 81.34\%\\
    \hline
    \multirow{4}{*}{CodeLlama-13b} & Instruct-hf& -  & 84.55\% & 93.78\%\\
    & CODEE&76k & 64.35\% & 76.36\%\\
    & BASIC&2k & 81.74\% & 89.33\%\\
    & MIX&2k & 71.93\% & 84.27\%\\
    \hline
    \end{tabular}
    }
    \vspace{-10pt}
    \label{tab:gpt4evaluation}
\end{table}

\finding{for \toolg}{The effectiveness of \toolg, validated through downstream
benchmarks (\S~\ref{subsubsec:main-results-bench}), internal logits analysis
(\S~\ref{subsubsec:logits-analysis}), and human evaluations
(\S~\ref{subsubsec:pairwise-comparison}), underscores the crucial role of APIs
in specific SFT dataset synthesis. By using APIs, \toolg\ frames specific SFT
dataset synthesis as a process of transforming high-level requirements into
concrete implementations, enabling the decomposition of complex generation
problems into manageable sub-problems. This approach allows \myframe\ to
robustly and efficiently synthesize high-quality data without relying on
real-world datasets or being constrained by specific powerful, proprietary
LLMs.}

\subsection{Alternative Generation Strategies}
\label{subsec:alternative-strategies}

An alternative approach to dataset generation draws inspiration from real-world
codebases. For instance, MagicCoder~\cite{weimagicoder} generates SFT datasets
using GitHub repository code snippets. To explore this method's potential in
domain-specific contexts, we conduct a comparison study by first replicating
MagicCoder's approach. We instruct models to reference original code and
incorporate five specific Numpy APIs (\texttt{np.squeeze()},
\texttt{np.random.uniform()}, \texttt{np.vstack()}, \texttt{np.var()}, and
\texttt{np.median()}) in their generated code.

\begin{figure}[!htbp]
    \centering
    \includegraphics[width=1.0\linewidth]{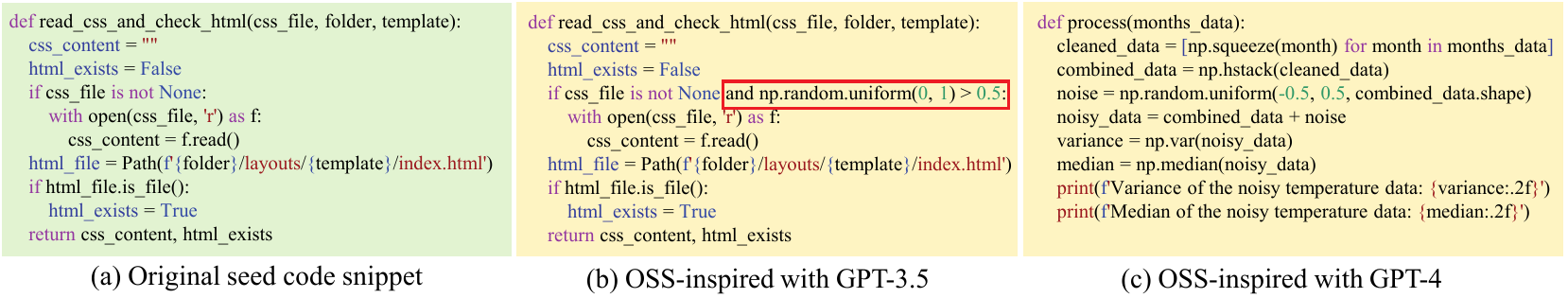}
    \vspace{-20pt}
    \caption{Example of alternative generation strategy with various models, showing only the generated code solutions. The accompanying problem descriptions are omitted for brevity.}
    \vspace{-10pt}
    \label{fig:magiccoder-35-example}
\end{figure}

\F~\ref{fig:magiccoder-35-example} illustrates the results of our test.
\F~\ref{fig:magiccoder-35-example}(a) shows the real-world code snippet used to inspire the LLM, while \F~\ref{fig:magiccoder-35-example}(b) displays the code generated by GPT-3.5. The output demonstrates GPT-3.5's difficulty in meeting the desired specifications. It merely adds a random condition \texttt{np.random.uniform(0, 1) > 0.5} to the existing \texttt{if} statement, failing to meaningfully utilize the Numpy APIs or alter the original code's semantics.
Furthermore, \F~\ref{fig:magiccoder-35-example}(c) showcases the output from GPT-4. While GPT-4 demonstrates improved performance by integrating more of the specified APIs, it still struggles to effectively draw inspiration from the provided code snippet. The resulting code lacks the complexity and sophisticated structure found in the original snippet.

This comparison study reveals potential limitations in the generation approach
based on real-world code snippets. When the model's capability is limited, it
may struggle to generate meaningful code. Even with more advanced models, the
generated code may lack the desired complexity. The insights gained from this
test highlight the strengths of \toolg's core design principles. By decomposing
complex problems to match LLM capabilities and specifying concrete APIs and code
structures, \toolg\ offers a more targeted and effective framework for
domain-specific dataset synthesis. This approach addresses the challenges
observed in the alternative strategy, potentially explaining \toolg's superior
performance in synthesizing high-quality, domain-specific datasets.

\section{Related Work}
\label{sec:related}

\parh{Component-based program synthesis.}~Program synthesis utilizing various levels of abstraction and specifications has demonstrated effectiveness in previous research~\cite{wang2017program,mell2024optimal,guria2023absynthe,feng2018program,polikarpova2016program}.
Our work specifically relates to program synthesis techniques that generate concise code fragments by utilizing components from existing libraries. The primary objective of these techniques is to support developers in programming tasks through library code reuse or to test complex software systems such as compilers. In these scenarios, a developer supplies an incomplete expression \cite{partial-expressions,insynth} or a method signature \cite{rest-synthesis,petri-net,type-refinement}, and the program synthesis tool generates a ranked list of implementation sketches that better align with the developer's intent. While these approaches strive to synthesize optimal solutions based on various factors (including code size, frequency of API method invocations, type distance, or user intent), our objective diverges. We aim to generate a diverse set of domain-specific programs to enhance SFT for LCMs. 

Our work draws inspiration from THALIA \cite{sotiropoulos2024api}, an API-driven program synthesis approach for testing the implementation of compilers' static typing procedures. However, THALIA emphasizes exploring the connections between different API implementations and associating them to synthesize meaningful programs that cover a wide range of type-related API usage patterns, aiding in compiler testing. In contrast, our method does not prioritize the meaningfulness of API combinations compared to existing code snippets. Instead, it focuses on creating unexpected programs that are rarely found in the existing codebase. Regarding implementation details, we reference SKETCH~\cite{solar2006combinatorial} in designing the ``hole'' concept in our code skeleton.
Our approach contributes to the SFT process by exposing the model to a more diverse set of programs, fostering a comprehensive understanding of the target library's API.

\parh{SFT Dataset Selection.}
LIMA~\cite{zhou2024lima} reveals that a limited amount of carefully curated human-written SFT data is sufficient to teach models to produce high-quality output. Furthermore, Dong et al.~\cite{dong2023abilities} discover that different abilities, such as math or reasoning, benefit from selecting varying proportions of data from the entire dataset. Based on these findings, numerous studies have focused on selecting the most representative data from the original dataset. Jiang et al.~\cite{jiang2024exploring} propose using learning complexity (LC) as a scoring function for data pruning in classification and regression tasks. Wang et al.~\cite{wang2024diversity} employ determinantal point processes to capture the diversity and quality of SFT datasets for subset selection, measuring dataset diversity with log determinant distance between the dataset of interest and a maximally diverse reference dataset. INSTRUCTMINING~\cite{cao2023instruction} combines customized language indicators with an advanced searching algorithm to automatically assess data quality and identify the optimal subset for fine-tuning language models. 

In addition to selecting the most representative data using indicator-based methods, some studies also consider employing DL models to aid in the selection process or using LLMs to rewrite the data for quality improvement. Ge et al.~\cite{ge2024clustering} propose Clustering and Ranking (CaR), a two-step approach that involves ranking instruction pairs using a scoring model aligned with expert preferences and preserving dataset diversity through a clustering process. CoachLM~\cite{liu2023automatic} uses an LLM fine-tuned on a coach finetuning dataset to rewrite the SFT data, enhancing the quality of fine-tuning datasets through automatic sample revisions.

\tools\ distinguishes itself from these methods in two key aspects. First, to the best of our knowledge, \tools\ is the first work specifically focusing on LCMs, while most existing work concentrates on natural language processing tasks and relies on natural language indicators for selection. Second, \tools\ is model-agnostic, meaning it does not require a deep learning model or training process information to aid in the selection process, unlike existing work that depends on such resources.

\parh{SFT Dataset Generation.}
Various methods have been developed to generate high-quality SFT datasets. Self-Instruct \cite{wang2023self} aligns LLMs with human intent using teacher LLM-generated data, while Evol-Instruct \cite{xu2023wizardlm} creates large volumes of instruction pairs step by step with varying complexity levels. These approaches have led to comprehensive datasets like Alpaca-GPT4 \cite{peng2023instruction} and Lmsys-chat-1m \cite{zheng2023lmsys}. Demonstrating the power of synthetic data, Nvidia's Nemotron-4 \cite{adler2024nemotron} achieves state-of-the-art performance using 98\% synthetically generated data in its alignment process.
For LCMs, Luo et al. \cite{WizardCoder} adapted Evol-Instruct to generate code snippets. Nemotron-4 further incorporated programming keywords in its generation process. However, these methods often lack diversity in seed programs. To address this, OSS-Instruct \cite{weimagicoder} integrates real-world code snippets, while AutoCoder \cite{lei2024autocoder} combines agent interaction with code execution verification to generate ``correct'' programs. 

\toolg\ distinguishes itself from existing methods in two key ways: First, it utilizes APIs in combination with program skeletons to generate high-quality SFT datasets, offering greater flexibility and data control than relying on real-world code snippets or specific topics. Second, \toolg's pipeline doesn't require any specific powerful model, further enhancing its capabilities and generalization.

\section{Conclusion}
\label{sec:conclusion}

We introduce \myframe, an API-guided dataset synthesis framework for enhancing
the fine-tuning process of LCMs. Our approach, comprising \tools\ for efficient
subset selection and \toolg\ for domain-specific dataset generation, addresses
the challenges of dataset quality and scarcity in both general and
domain-specific scenarios. Extensive experiments demonstrate the effectiveness
of our framework, with models fine-tuned on datasets constructed using \tools\
and \toolg\ outperforming those tuned on larger, unoptimized datasets. Our work
highlights the crucial role of APIs in guiding the synthesis of high-quality
datasets for LCM fine-tuning. By leveraging API-level abstractions, we offer a
novel perspective on dataset synthesis that improves the efficiency and
effectiveness of fine-tuning. This API-guided approach to dataset synthesis not
only enhances model performance but also provides a scalable solution for both
general and domain-specific applications of LCMs.
We believe that further
research in this direction will lead to more powerful and adaptable LCMs,
opening new avenues for AI-assisted software development.

\bibliographystyle{ACM-Reference-Format}
\bibliography{./bib/code, ./bib/prosyn,./bib/sft,./bib/cot}


\begin{thebibliography}{90}


\ifx \showCODEN    \undefined \def \showCODEN     #1{\unskip}     \fi
\ifx \showDOI      \undefined \def \showDOI       #1{#1}\fi
\ifx \showISBNx    \undefined \def \showISBNx     #1{\unskip}     \fi
\ifx \showISBNxiii \undefined \def \showISBNxiii  #1{\unskip}     \fi
\ifx \showISSN     \undefined \def \showISSN      #1{\unskip}     \fi
\ifx \showLCCN     \undefined \def \showLCCN      #1{\unskip}     \fi
\ifx \shownote     \undefined \def \shownote      #1{#1}          \fi
\ifx \showarticletitle \undefined \def \showarticletitle #1{#1}   \fi
\ifx \showURL      \undefined \def \showURL       {\relax}        \fi
\providecommand\bibfield[2]{#2}
\providecommand\bibinfo[2]{#2}
\providecommand\natexlab[1]{#1}
\providecommand\showeprint[2][]{arXiv:#2}

\bibitem[Abdi and Williams(2010)]%
        {abdi2010principal}
\bibfield{author}{\bibinfo{person}{Herv{\'e} Abdi} {and}
  \bibinfo{person}{Lynne~J Williams}.} \bibinfo{year}{2010}\natexlab{}.
\newblock \showarticletitle{Principal component analysis}.
\newblock \bibinfo{journal}{\emph{Wiley interdisciplinary reviews:
  computational statistics}} \bibinfo{volume}{2}, \bibinfo{number}{4}
  (\bibinfo{year}{2010}), \bibinfo{pages}{433--459}.
\newblock


\bibitem[Abdin et~al\mbox{.}(2024)]%
        {abdin2024phi}
\bibfield{author}{\bibinfo{person}{Marah Abdin}, \bibinfo{person}{Sam~Ade
  Jacobs}, \bibinfo{person}{Ammar~Ahmad Awan}, \bibinfo{person}{Jyoti Aneja},
  \bibinfo{person}{Ahmed Awadallah}, \bibinfo{person}{Hany Awadalla},
  \bibinfo{person}{Nguyen Bach}, \bibinfo{person}{Amit Bahree},
  \bibinfo{person}{Arash Bakhtiari}, \bibinfo{person}{Harkirat Behl},
  {et~al\mbox{.}}} \bibinfo{year}{2024}\natexlab{}.
\newblock \showarticletitle{Phi-3 technical report: A highly capable language
  model locally on your phone}.
\newblock \bibinfo{journal}{\emph{arXiv preprint arXiv:2404.14219}}
  (\bibinfo{year}{2024}).
\newblock


\bibitem[Adler et~al\mbox{.}(2024)]%
        {adler2024nemotron}
\bibfield{author}{\bibinfo{person}{Bo Adler}, \bibinfo{person}{Niket Agarwal},
  \bibinfo{person}{Ashwath Aithal}, \bibinfo{person}{Dong~H Anh},
  \bibinfo{person}{Pallab Bhattacharya}, \bibinfo{person}{Annika Brundyn},
  \bibinfo{person}{Jared Casper}, \bibinfo{person}{Bryan Catanzaro},
  \bibinfo{person}{Sharon Clay}, \bibinfo{person}{Jonathan Cohen},
  {et~al\mbox{.}}} \bibinfo{year}{2024}\natexlab{}.
\newblock \showarticletitle{Nemotron-4 340B Technical Report}.
\newblock \bibinfo{journal}{\emph{arXiv preprint arXiv:2406.11704}}
  (\bibinfo{year}{2024}).
\newblock


\bibitem[Al-Kaswan et~al\mbox{.}(2024)]%
        {al2024traces}
\bibfield{author}{\bibinfo{person}{Ali Al-Kaswan}, \bibinfo{person}{Maliheh
  Izadi}, {and} \bibinfo{person}{Arie Van~Deursen}.}
  \bibinfo{year}{2024}\natexlab{}.
\newblock \showarticletitle{Traces of memorisation in large language models for
  code}. In \bibinfo{booktitle}{\emph{Proceedings of the IEEE/ACM 46th
  International Conference on Software Engineering}}. \bibinfo{pages}{1--12}.
\newblock


\bibitem[Bai et~al\mbox{.}(2023)]%
        {bai2023qwen}
\bibfield{author}{\bibinfo{person}{Jinze Bai}, \bibinfo{person}{Shuai Bai},
  \bibinfo{person}{Yunfei Chu}, \bibinfo{person}{Zeyu Cui},
  \bibinfo{person}{Kai Dang}, \bibinfo{person}{Xiaodong Deng},
  \bibinfo{person}{Yang Fan}, \bibinfo{person}{Wenbin Ge}, \bibinfo{person}{Yu
  Han}, \bibinfo{person}{Fei Huang}, {et~al\mbox{.}}}
  \bibinfo{year}{2023}\natexlab{}.
\newblock \showarticletitle{Qwen technical report}.
\newblock \bibinfo{journal}{\emph{arXiv preprint arXiv:2309.16609}}
  (\bibinfo{year}{2023}).
\newblock


\bibitem[Brown et~al\mbox{.}(2020)]%
        {brown2020language}
\bibfield{author}{\bibinfo{person}{Tom Brown}, \bibinfo{person}{Benjamin Mann},
  \bibinfo{person}{Nick Ryder}, \bibinfo{person}{Melanie Subbiah},
  \bibinfo{person}{Jared~D Kaplan}, \bibinfo{person}{Prafulla Dhariwal},
  \bibinfo{person}{Arvind Neelakantan}, \bibinfo{person}{Pranav Shyam},
  \bibinfo{person}{Girish Sastry}, \bibinfo{person}{Amanda Askell},
  {et~al\mbox{.}}} \bibinfo{year}{2020}\natexlab{}.
\newblock \showarticletitle{Language models are few-shot learners}.
\newblock \bibinfo{journal}{\emph{Advances in neural information processing
  systems}}  \bibinfo{volume}{33} (\bibinfo{year}{2020}),
  \bibinfo{pages}{1877--1901}.
\newblock


\bibitem[Burns et~al\mbox{.}(2023)]%
        {burns2023weak}
\bibfield{author}{\bibinfo{person}{Collin Burns}, \bibinfo{person}{Pavel
  Izmailov}, \bibinfo{person}{Jan~Hendrik Kirchner}, \bibinfo{person}{Bowen
  Baker}, \bibinfo{person}{Leo Gao}, \bibinfo{person}{Leopold Aschenbrenner},
  \bibinfo{person}{Yining Chen}, \bibinfo{person}{Adrien Ecoffet},
  \bibinfo{person}{Manas Joglekar}, \bibinfo{person}{Jan Leike},
  {et~al\mbox{.}}} \bibinfo{year}{2023}\natexlab{}.
\newblock \showarticletitle{Weak-to-strong generalization: Eliciting strong
  capabilities with weak supervision}.
\newblock \bibinfo{journal}{\emph{arXiv preprint arXiv:2312.09390}}
  (\bibinfo{year}{2023}).
\newblock


\bibitem[Cao et~al\mbox{.}(2023)]%
        {cao2023instruction}
\bibfield{author}{\bibinfo{person}{Yihan Cao}, \bibinfo{person}{Yanbin Kang},
  \bibinfo{person}{Chi Wang}, {and} \bibinfo{person}{Lichao Sun}.}
  \bibinfo{year}{2023}\natexlab{}.
\newblock \showarticletitle{Instruction mining: When data mining meets large
  language model finetuning}.
\newblock \bibinfo{journal}{\emph{arXiv preprint arXiv}}
  \bibinfo{volume}{2307} (\bibinfo{year}{2023}).
\newblock


\bibitem[Chen et~al\mbox{.}(2021)]%
        {chen2021evaluating}
\bibfield{author}{\bibinfo{person}{Mark Chen}, \bibinfo{person}{Jerry Tworek},
  \bibinfo{person}{Heewoo Jun}, \bibinfo{person}{Qiming Yuan},
  \bibinfo{person}{Henrique Ponde de~Oliveira Pinto}, \bibinfo{person}{Jared
  Kaplan}, \bibinfo{person}{Harri Edwards}, \bibinfo{person}{Yuri Burda},
  \bibinfo{person}{Nicholas Joseph}, \bibinfo{person}{Greg Brockman},
  {et~al\mbox{.}}} \bibinfo{year}{2021}\natexlab{}.
\newblock \showarticletitle{Evaluating large language models trained on code}.
\newblock \bibinfo{journal}{\emph{arXiv preprint arXiv:2107.03374}}
  (\bibinfo{year}{2021}).
\newblock


\bibitem[Cheng et~al\mbox{.}(2024)]%
        {cheng2024adapting}
\bibfield{author}{\bibinfo{person}{Daixuan Cheng}, \bibinfo{person}{Shaohan
  Huang}, {and} \bibinfo{person}{Furu Wei}.} \bibinfo{year}{2024}\natexlab{}.
\newblock \showarticletitle{Adapting Large Language Models via Reading
  Comprehension}. In \bibinfo{booktitle}{\emph{The Twelfth International
  Conference on Learning Representations}}.
\newblock
\urldef\tempurl%
\url{https://openreview.net/forum?id=y886UXPEZ0}
\showURL{%
\tempurl}


\bibitem[Chung et~al\mbox{.}(2024)]%
        {chung2024scaling}
\bibfield{author}{\bibinfo{person}{Hyung~Won Chung}, \bibinfo{person}{Le Hou},
  \bibinfo{person}{Shayne Longpre}, \bibinfo{person}{Barret Zoph},
  \bibinfo{person}{Yi Tay}, \bibinfo{person}{William Fedus},
  \bibinfo{person}{Yunxuan Li}, \bibinfo{person}{Xuezhi Wang},
  \bibinfo{person}{Mostafa Dehghani}, \bibinfo{person}{Siddhartha Brahma},
  {et~al\mbox{.}}} \bibinfo{year}{2024}\natexlab{}.
\newblock \showarticletitle{Scaling instruction-finetuned language models}.
\newblock \bibinfo{journal}{\emph{Journal of Machine Learning Research}}
  \bibinfo{volume}{25}, \bibinfo{number}{70} (\bibinfo{year}{2024}),
  \bibinfo{pages}{1--53}.
\newblock


\bibitem[{codefuse ai}(2023)]%
        {CodeExercise}
\bibfield{author}{\bibinfo{person}{{codefuse ai}}.}
  \bibinfo{year}{2023}\natexlab{}.
\newblock \bibinfo{title}{{CodeExercise-Python-27k}}.
\newblock
\newblock
\urldef\tempurl%
\url{https://huggingface.co/datasets/codefuse-ai/CodeExercise-Python-27k/}
\showURL{%
\tempurl}


\bibitem[Daniel et~al\mbox{.}(2018)]%
        {daniel2018quality}
\bibfield{author}{\bibinfo{person}{Florian Daniel}, \bibinfo{person}{Pavel
  Kucherbaev}, \bibinfo{person}{Cinzia Cappiello}, \bibinfo{person}{Boualem
  Benatallah}, {and} \bibinfo{person}{Mohammad Allahbakhsh}.}
  \bibinfo{year}{2018}\natexlab{}.
\newblock \showarticletitle{Quality control in crowdsourcing: A survey of
  quality attributes, assessment techniques, and assurance actions}.
\newblock \bibinfo{journal}{\emph{ACM Computing Surveys (CSUR)}}
  \bibinfo{volume}{51}, \bibinfo{number}{1} (\bibinfo{year}{2018}),
  \bibinfo{pages}{1--40}.
\newblock


\bibitem[{Defog AI}(2024)]%
        {Sqlcoder2}
\bibfield{author}{\bibinfo{person}{{Defog AI}}.}
  \bibinfo{year}{2024}\natexlab{}.
\newblock \bibinfo{title}{{Open-sourcing SQLCoder2-15b and SQLCoder-7b}}.
\newblock
\newblock
\urldef\tempurl%
\url{https://defog.ai/blog/open-sourcing-sqlcoder2-7b/}
\showURL{%
\tempurl}


\bibitem[Ding et~al\mbox{.}(2024)]%
        {ding2024cycle}
\bibfield{author}{\bibinfo{person}{Yangruibo Ding}, \bibinfo{person}{Marcus~J
  Min}, \bibinfo{person}{Gail Kaiser}, {and} \bibinfo{person}{Baishakhi Ray}.}
  \bibinfo{year}{2024}\natexlab{}.
\newblock \showarticletitle{Cycle: Learning to self-refine the code
  generation}.
\newblock \bibinfo{journal}{\emph{Proceedings of the ACM on Programming
  Languages}} \bibinfo{volume}{8}, \bibinfo{number}{OOPSLA1}
  (\bibinfo{year}{2024}), \bibinfo{pages}{392--418}.
\newblock


\bibitem[Dong et~al\mbox{.}(2023)]%
        {dong2023abilities}
\bibfield{author}{\bibinfo{person}{Guanting Dong}, \bibinfo{person}{Hongyi
  Yuan}, \bibinfo{person}{Keming Lu}, \bibinfo{person}{Chengpeng Li},
  \bibinfo{person}{Mingfeng Xue}, \bibinfo{person}{Dayiheng Liu},
  \bibinfo{person}{Wei Wang}, \bibinfo{person}{Zheng Yuan},
  \bibinfo{person}{Chang Zhou}, {and} \bibinfo{person}{Jingren Zhou}.}
  \bibinfo{year}{2023}\natexlab{}.
\newblock \showarticletitle{How abilities in large language models are affected
  by supervised fine-tuning data composition}.
\newblock \bibinfo{journal}{\emph{arXiv preprint arXiv:2310.05492}}
  (\bibinfo{year}{2023}).
\newblock


\bibitem[Feng et~al\mbox{.}(2018)]%
        {feng2018program}
\bibfield{author}{\bibinfo{person}{Yu Feng}, \bibinfo{person}{Ruben Martins},
  \bibinfo{person}{Osbert Bastani}, {and} \bibinfo{person}{Isil Dillig}.}
  \bibinfo{year}{2018}\natexlab{}.
\newblock \showarticletitle{Program synthesis using conflict-driven learning}.
\newblock \bibinfo{journal}{\emph{ACM SIGPLAN Notices}} \bibinfo{volume}{53},
  \bibinfo{number}{4} (\bibinfo{year}{2018}), \bibinfo{pages}{420--435}.
\newblock


\bibitem[Feng et~al\mbox{.}(2017)]%
        {petri-net}
\bibfield{author}{\bibinfo{person}{Yu Feng}, \bibinfo{person}{Ruben Martins},
  \bibinfo{person}{Yuepeng Wang}, \bibinfo{person}{Isil Dillig}, {and}
  \bibinfo{person}{Thomas~W. Reps}.} \bibinfo{year}{2017}\natexlab{}.
\newblock \showarticletitle{Component-Based Synthesis for Complex {API}s}
  \emph{(\bibinfo{series}{POPL '17})}. \bibinfo{publisher}{Association for
  Computing Machinery}, \bibinfo{address}{New York, NY, USA},
  \bibinfo{pages}{599–612}.
\newblock
\showISBNx{9781450346603}
\urldef\tempurl%
\url{https://doi.org/10.1145/3009837.3009851}
\showDOI{\tempurl}


\bibitem[Fleiss(1971)]%
        {fleiss1971measuring}
\bibfield{author}{\bibinfo{person}{Joseph~L Fleiss}.}
  \bibinfo{year}{1971}\natexlab{}.
\newblock \showarticletitle{Measuring nominal scale agreement among many
  raters.}
\newblock \bibinfo{journal}{\emph{Psychological bulletin}}
  \bibinfo{volume}{76}, \bibinfo{number}{5} (\bibinfo{year}{1971}),
  \bibinfo{pages}{378}.
\newblock


\bibitem[Fraser and Arcuri(2011)]%
        {fraser2011evosuite}
\bibfield{author}{\bibinfo{person}{Gordon Fraser} {and} \bibinfo{person}{Andrea
  Arcuri}.} \bibinfo{year}{2011}\natexlab{}.
\newblock \showarticletitle{Evosuite: automatic test suite generation for
  object-oriented software}. In \bibinfo{booktitle}{\emph{Proceedings of the
  19th ACM SIGSOFT symposium and the 13th European conference on Foundations of
  software engineering}}. \bibinfo{pages}{416--419}.
\newblock


\bibitem[Fuglede and Topsoe(2004)]%
        {fuglede2004jensen}
\bibfield{author}{\bibinfo{person}{Bent Fuglede} {and}
  \bibinfo{person}{Flemming Topsoe}.} \bibinfo{year}{2004}\natexlab{}.
\newblock \showarticletitle{Jensen-Shannon divergence and Hilbert space
  embedding}. In \bibinfo{booktitle}{\emph{International symposium
  onInformation theory, 2004. ISIT 2004. Proceedings.}} IEEE,
  \bibinfo{pages}{31}.
\newblock


\bibitem[Ge et~al\mbox{.}(2024)]%
        {ge2024clustering}
\bibfield{author}{\bibinfo{person}{Yuan Ge}, \bibinfo{person}{Yilun Liu},
  \bibinfo{person}{Chi Hu}, \bibinfo{person}{Weibin Meng},
  \bibinfo{person}{Shimin Tao}, \bibinfo{person}{Xiaofeng Zhao},
  \bibinfo{person}{Hongxia Ma}, \bibinfo{person}{Li Zhang},
  \bibinfo{person}{Hao Yang}, {and} \bibinfo{person}{Tong Xiao}.}
  \bibinfo{year}{2024}\natexlab{}.
\newblock \showarticletitle{Clustering and Ranking: Diversity-preserved
  Instruction Selection through Expert-aligned Quality Estimation}.
\newblock \bibinfo{journal}{\emph{arXiv preprint arXiv:2402.18191}}
  (\bibinfo{year}{2024}).
\newblock


\bibitem[Ghosh et~al\mbox{.}(2024)]%
        {ghosh2024closer}
\bibfield{author}{\bibinfo{person}{Sreyan Ghosh}, \bibinfo{person}{Chandra
  Kiran~Reddy Evuru}, \bibinfo{person}{Sonal Kumar}, \bibinfo{person}{Deepali
  Aneja}, \bibinfo{person}{Zeyu Jin}, \bibinfo{person}{Ramani Duraiswami},
  \bibinfo{person}{Dinesh Manocha}, {et~al\mbox{.}}}
  \bibinfo{year}{2024}\natexlab{}.
\newblock \showarticletitle{A Closer Look at the Limitations of Instruction
  Tuning}.
\newblock \bibinfo{journal}{\emph{arXiv preprint arXiv:2402.05119}}
  (\bibinfo{year}{2024}).
\newblock


\bibitem[Glorot and Bengio(2010)]%
        {glorot2010understanding}
\bibfield{author}{\bibinfo{person}{Xavier Glorot} {and} \bibinfo{person}{Yoshua
  Bengio}.} \bibinfo{year}{2010}\natexlab{}.
\newblock \showarticletitle{Understanding the difficulty of training deep
  feedforward neural networks}. In \bibinfo{booktitle}{\emph{Proceedings of the
  thirteenth international conference on artificial intelligence and
  statistics}}. JMLR Workshop and Conference Proceedings,
  \bibinfo{pages}{249--256}.
\newblock


\bibitem[{Google Cloud}(2023)]%
        {secpalm}
\bibfield{author}{\bibinfo{person}{{Google Cloud}}.}
  \bibinfo{year}{2023}\natexlab{}.
\newblock \bibinfo{title}{{Supercharging security with generative AI}}.
\newblock
\newblock
\urldef\tempurl%
\url{https://cloud.google.com/blog/products/identity-security/rsa-google-cloud-security-ai-workbench-generative-ai}
\showURL{%
\tempurl}


\bibitem[Guo et~al\mbox{.}(2022)]%
        {rest-synthesis}
\bibfield{author}{\bibinfo{person}{Zheng Guo}, \bibinfo{person}{David Cao},
  \bibinfo{person}{Davin Tjong}, \bibinfo{person}{Jean Yang},
  \bibinfo{person}{Cole Schlesinger}, {and} \bibinfo{person}{Nadia
  Polikarpova}.} \bibinfo{year}{2022}\natexlab{}.
\newblock \showarticletitle{Type-Directed Program Synthesis for {REST}ful
  {API}s}. In \bibinfo{booktitle}{\emph{Proceedings of the 43rd ACM SIGPLAN
  International Conference on Programming Language Design and Implementation}}
  (San Diego, CA, USA) \emph{(\bibinfo{series}{PLDI 2022})}.
  \bibinfo{publisher}{Association for Computing Machinery},
  \bibinfo{address}{New York, NY, USA}, \bibinfo{pages}{122–136}.
\newblock
\showISBNx{9781450392655}
\urldef\tempurl%
\url{https://doi.org/10.1145/3519939.3523450}
\showDOI{\tempurl}


\bibitem[Guo et~al\mbox{.}(2019)]%
        {type-refinement}
\bibfield{author}{\bibinfo{person}{Zheng Guo}, \bibinfo{person}{Michael James},
  \bibinfo{person}{David Justo}, \bibinfo{person}{Jiaxiao Zhou},
  \bibinfo{person}{Ziteng Wang}, \bibinfo{person}{Ranjit Jhala}, {and}
  \bibinfo{person}{Nadia Polikarpova}.} \bibinfo{year}{2019}\natexlab{}.
\newblock \showarticletitle{Program Synthesis by Type-Guided Abstraction
  Refinement}.
\newblock \bibinfo{journal}{\emph{Proc. ACM Program. Lang.}}
  \bibinfo{volume}{4}, \bibinfo{number}{POPL}, Article \bibinfo{articleno}{12}
  (\bibinfo{date}{dec} \bibinfo{year}{2019}), \bibinfo{numpages}{28}~pages.
\newblock
\urldef\tempurl%
\url{https://doi.org/10.1145/3371080}
\showDOI{\tempurl}


\bibitem[Guria et~al\mbox{.}(2023)]%
        {guria2023absynthe}
\bibfield{author}{\bibinfo{person}{Sankha~Narayan Guria},
  \bibinfo{person}{Jeffrey~S Foster}, {and} \bibinfo{person}{David Van~Horn}.}
  \bibinfo{year}{2023}\natexlab{}.
\newblock \showarticletitle{Absynthe: Abstract Interpretation-Guided
  Synthesis}.
\newblock \bibinfo{journal}{\emph{Proceedings of the ACM on Programming
  Languages}} \bibinfo{volume}{7}, \bibinfo{number}{PLDI}
  (\bibinfo{year}{2023}), \bibinfo{pages}{1584--1607}.
\newblock


\bibitem[Gvero et~al\mbox{.}(2013)]%
        {insynth}
\bibfield{author}{\bibinfo{person}{Tihomir Gvero}, \bibinfo{person}{Viktor
  Kuncak}, \bibinfo{person}{Ivan Kuraj}, {and} \bibinfo{person}{Ruzica
  Piskac}.} \bibinfo{year}{2013}\natexlab{}.
\newblock \showarticletitle{Complete Completion Using Types and Weights}
  \emph{(\bibinfo{series}{PLDI '13})}. \bibinfo{publisher}{Association for
  Computing Machinery}, \bibinfo{address}{New York, NY, USA},
  \bibinfo{pages}{27–38}.
\newblock
\showISBNx{9781450320146}
\urldef\tempurl%
\url{https://doi.org/10.1145/2491956.2462192}
\showDOI{\tempurl}


\bibitem[Han et~al\mbox{.}(2024)]%
        {han2024parameter}
\bibfield{author}{\bibinfo{person}{Zeyu Han}, \bibinfo{person}{Chao Gao},
  \bibinfo{person}{Jinyang Liu}, \bibinfo{person}{Sai~Qian Zhang},
  {et~al\mbox{.}}} \bibinfo{year}{2024}\natexlab{}.
\newblock \showarticletitle{Parameter-efficient fine-tuning for large models: A
  comprehensive survey}.
\newblock \bibinfo{journal}{\emph{arXiv preprint arXiv:2403.14608}}
  (\bibinfo{year}{2024}).
\newblock


\bibitem[Hartigan and Wong(1979)]%
        {kmeans1979algorithm}
\bibfield{author}{\bibinfo{person}{John~A Hartigan} {and}
  \bibinfo{person}{Manchek~A Wong}.} \bibinfo{year}{1979}\natexlab{}.
\newblock \showarticletitle{Algorithm AS 136: A k-means clustering algorithm}.
\newblock \bibinfo{journal}{\emph{Journal of the royal statistical society.
  series c (applied statistics)}} \bibinfo{volume}{28}, \bibinfo{number}{1}
  (\bibinfo{year}{1979}), \bibinfo{pages}{100--108}.
\newblock


\bibitem[Jiang et~al\mbox{.}(2024)]%
        {jiang2024exploring}
\bibfield{author}{\bibinfo{person}{Wenyu Jiang}, \bibinfo{person}{Zhenlong
  Liu}, \bibinfo{person}{Zejian Xie}, \bibinfo{person}{Songxin Zhang},
  \bibinfo{person}{Bingyi Jing}, {and} \bibinfo{person}{Hongxin Wei}.}
  \bibinfo{year}{2024}\natexlab{}.
\newblock \showarticletitle{Exploring Learning Complexity for Downstream Data
  Pruning}.
\newblock \bibinfo{journal}{\emph{arXiv preprint arXiv:2402.05356}}
  (\bibinfo{year}{2024}).
\newblock


\bibitem[Kingma and Ba(2015)]%
        {KingBa15}
\bibfield{author}{\bibinfo{person}{Diederik Kingma} {and}
  \bibinfo{person}{Jimmy Ba}.} \bibinfo{year}{2015}\natexlab{}.
\newblock \showarticletitle{Adam: A Method for Stochastic Optimization}. In
  \bibinfo{booktitle}{\emph{International Conference on Learning
  Representations (ICLR)}}. \bibinfo{address}{San Diega, CA, USA}.
\newblock


\bibitem[Kojima et~al\mbox{.}(2022)]%
        {kojima2022large}
\bibfield{author}{\bibinfo{person}{Takeshi Kojima},
  \bibinfo{person}{Shixiang~Shane Gu}, \bibinfo{person}{Machel Reid},
  \bibinfo{person}{Yutaka Matsuo}, {and} \bibinfo{person}{Yusuke Iwasawa}.}
  \bibinfo{year}{2022}\natexlab{}.
\newblock \showarticletitle{Large language models are zero-shot reasoners}.
\newblock \bibinfo{journal}{\emph{Advances in neural information processing
  systems}}  \bibinfo{volume}{35} (\bibinfo{year}{2022}),
  \bibinfo{pages}{22199--22213}.
\newblock


\bibitem[K{\"o}pf et~al\mbox{.}(2024)]%
        {kopf2024openassistant}
\bibfield{author}{\bibinfo{person}{Andreas K{\"o}pf}, \bibinfo{person}{Yannic
  Kilcher}, \bibinfo{person}{Dimitri von R{\"u}tte}, \bibinfo{person}{Sotiris
  Anagnostidis}, \bibinfo{person}{Zhi~Rui Tam}, \bibinfo{person}{Keith
  Stevens}, \bibinfo{person}{Abdullah Barhoum}, \bibinfo{person}{Duc Nguyen},
  \bibinfo{person}{Oliver Stanley}, \bibinfo{person}{Rich{\'a}rd Nagyfi},
  {et~al\mbox{.}}} \bibinfo{year}{2024}\natexlab{}.
\newblock \showarticletitle{Openassistant conversations-democratizing large
  language model alignment}.
\newblock \bibinfo{journal}{\emph{Advances in Neural Information Processing
  Systems}}  \bibinfo{volume}{36} (\bibinfo{year}{2024}).
\newblock


\bibitem[Lei et~al\mbox{.}(2024)]%
        {lei2024autocoder}
\bibfield{author}{\bibinfo{person}{Bin Lei}, \bibinfo{person}{Yuchen Li}, {and}
  \bibinfo{person}{Qiuwu Chen}.} \bibinfo{year}{2024}\natexlab{}.
\newblock \showarticletitle{AutoCoder: Enhancing Code Large Language Model
  with$\backslash$textsc $\{$AIEV-Instruct$\}$}.
\newblock \bibinfo{journal}{\emph{arXiv preprint arXiv:2405.14906}}
  (\bibinfo{year}{2024}).
\newblock


\bibitem[Li et~al\mbox{.}(2023a)]%
        {li2023starcoder}
\bibfield{author}{\bibinfo{person}{Raymond Li}, \bibinfo{person}{Loubna~Ben
  Allal}, \bibinfo{person}{Yangtian Zi}, \bibinfo{person}{Niklas Muennighoff},
  \bibinfo{person}{Denis Kocetkov}, \bibinfo{person}{Chenghao Mou},
  \bibinfo{person}{Marc Marone}, \bibinfo{person}{Christopher Akiki},
  \bibinfo{person}{Jia Li}, \bibinfo{person}{Jenny Chim}, {et~al\mbox{.}}}
  \bibinfo{year}{2023}\natexlab{a}.
\newblock \showarticletitle{Starcoder: may the source be with you!}
\newblock \bibinfo{journal}{\emph{arXiv preprint arXiv:2305.06161}}
  (\bibinfo{year}{2023}).
\newblock


\bibitem[Li et~al\mbox{.}(2023b)]%
        {StarCoder}
\bibfield{author}{\bibinfo{person}{Raymond Li}, \bibinfo{person}{Loubna~Ben
  Allal}, \bibinfo{person}{Yangtian Zi}, \bibinfo{person}{Niklas Muennighoff},
  \bibinfo{person}{Denis Kocetkov}, \bibinfo{person}{Chenghao Mou},
  \bibinfo{person}{Marc Marone}, \bibinfo{person}{Christopher Akiki},
  \bibinfo{person}{Jia Li}, \bibinfo{person}{Jenny Chim}, {et~al\mbox{.}}}
  \bibinfo{year}{2023}\natexlab{b}.
\newblock \showarticletitle{StarCoder: may the source be with you!}
\newblock \bibinfo{journal}{\emph{arXiv preprint arXiv:2305.06161}}
  (\bibinfo{year}{2023}).
\newblock


\bibitem[Liu et~al\mbox{.}(2023)]%
        {liu2023automatic}
\bibfield{author}{\bibinfo{person}{Yilun Liu}, \bibinfo{person}{Shimin Tao},
  \bibinfo{person}{Xiaofeng Zhao}, \bibinfo{person}{Ming Zhu},
  \bibinfo{person}{Wenbing Ma}, \bibinfo{person}{Junhao Zhu},
  \bibinfo{person}{Chang Su}, \bibinfo{person}{Yutai Hou},
  \bibinfo{person}{Miao Zhang}, \bibinfo{person}{Min Zhang}, {et~al\mbox{.}}}
  \bibinfo{year}{2023}\natexlab{}.
\newblock \showarticletitle{Automatic instruction optimization for open-source
  llm instruction tuning}.
\newblock \bibinfo{journal}{\emph{arXiv preprint arXiv:2311.13246}}
  (\bibinfo{year}{2023}).
\newblock


\bibitem[Liu et~al\mbox{.}(2024)]%
        {liu2024coachlm}
\bibfield{author}{\bibinfo{person}{Yilun Liu}, \bibinfo{person}{Shimin Tao},
  \bibinfo{person}{Xiaofeng Zhao}, \bibinfo{person}{Ming Zhu},
  \bibinfo{person}{Wenbing Ma}, \bibinfo{person}{Junhao Zhu},
  \bibinfo{person}{Chang Su}, \bibinfo{person}{Yutai Hou},
  \bibinfo{person}{Miao Zhang}, \bibinfo{person}{Min Zhang},
  \bibinfo{person}{Hongxia Ma}, \bibinfo{person}{Li Zhang},
  \bibinfo{person}{Hao Yang}, {and} \bibinfo{person}{Yanfei Jiang}.}
  \bibinfo{year}{2024}\natexlab{}.
\newblock \bibinfo{title}{CoachLM: Automatic Instruction Revisions Improve the
  Data Quality in LLM Instruction Tuning}.
\newblock
\newblock
\showeprint[arxiv]{2311.13246}


\bibitem[Loshchilov and Hutter(2017)]%
        {loshchilov2017sgdr}
\bibfield{author}{\bibinfo{person}{Ilya Loshchilov} {and}
  \bibinfo{person}{Frank Hutter}.} \bibinfo{year}{2017}\natexlab{}.
\newblock \showarticletitle{{SGDR}: Stochastic Gradient Descent with Warm
  Restarts}. In \bibinfo{booktitle}{\emph{International Conference on Learning
  Representations}}.
\newblock
\urldef\tempurl%
\url{https://openreview.net/forum?id=Skq89Scxx}
\showURL{%
\tempurl}


\bibitem[Lou et~al\mbox{.}(2015)]%
        {lou2015mutation}
\bibfield{author}{\bibinfo{person}{Yiling Lou}, \bibinfo{person}{Dan Hao},
  {and} \bibinfo{person}{Lu Zhang}.} \bibinfo{year}{2015}\natexlab{}.
\newblock \showarticletitle{Mutation-based test-case prioritization in software
  evolution}. In \bibinfo{booktitle}{\emph{2015 IEEE 26th International
  Symposium on Software Reliability Engineering (ISSRE)}}. IEEE,
  \bibinfo{pages}{46--57}.
\newblock


\bibitem[Lu et~al\mbox{.}(2021)]%
        {lu2021codexglue}
\bibfield{author}{\bibinfo{person}{Shuai Lu}, \bibinfo{person}{Daya Guo},
  \bibinfo{person}{Shuo Ren}, \bibinfo{person}{Junjie Huang},
  \bibinfo{person}{Alexey Svyatkovskiy}, \bibinfo{person}{Ambrosio Blanco},
  \bibinfo{person}{Colin~B. Clement}, \bibinfo{person}{Dawn Drain},
  \bibinfo{person}{Daxin Jiang}, \bibinfo{person}{Duyu Tang},
  \bibinfo{person}{Ge Li}, \bibinfo{person}{Lidong Zhou},
  \bibinfo{person}{Linjun Shou}, \bibinfo{person}{Long Zhou},
  \bibinfo{person}{Michele Tufano}, \bibinfo{person}{Ming Gong},
  \bibinfo{person}{Ming Zhou}, \bibinfo{person}{Nan Duan},
  \bibinfo{person}{Neel Sundaresan}, \bibinfo{person}{Shao~Kun Deng},
  \bibinfo{person}{Shengyu Fu}, {and} \bibinfo{person}{Shujie Liu}.}
  \bibinfo{year}{2021}\natexlab{}.
\newblock \showarticletitle{CodeXGLUE: {A} Machine Learning Benchmark Dataset
  for Code Understanding and Generation}. In \bibinfo{booktitle}{\emph{NeurIPS
  Datasets and Benchmarks 2021, virtual}},
  \bibfield{editor}{\bibinfo{person}{Joaquin Vanschoren} {and}
  \bibinfo{person}{Sai{-}Kit Yeung}} (Eds.).
\newblock


\bibitem[Luo et~al\mbox{.}(2023a)]%
        {WizardCoder}
\bibfield{author}{\bibinfo{person}{Ziyang Luo}, \bibinfo{person}{Can Xu},
  \bibinfo{person}{Pu Zhao}, \bibinfo{person}{Qingfeng Sun},
  \bibinfo{person}{Xiubo Geng}, \bibinfo{person}{Wenxiang Hu},
  \bibinfo{person}{Chongyang Tao}, \bibinfo{person}{Jing Ma},
  \bibinfo{person}{Qingwei Lin}, {and} \bibinfo{person}{Daxin Jiang}.}
  \bibinfo{year}{2023}\natexlab{a}.
\newblock \showarticletitle{WizardCoder: Empowering Code Large Language Models
  with Evol-Instruct}.
\newblock \bibinfo{journal}{\emph{CoRR}}  \bibinfo{volume}{abs/2306.08568}
  (\bibinfo{year}{2023}).
\newblock


\bibitem[Luo et~al\mbox{.}(2023b)]%
        {luo2023wizardcoder}
\bibfield{author}{\bibinfo{person}{Ziyang Luo}, \bibinfo{person}{Can Xu},
  \bibinfo{person}{Pu Zhao}, \bibinfo{person}{Qingfeng Sun},
  \bibinfo{person}{Xiubo Geng}, \bibinfo{person}{Wenxiang Hu},
  \bibinfo{person}{Chongyang Tao}, \bibinfo{person}{Jing Ma},
  \bibinfo{person}{Qingwei Lin}, {and} \bibinfo{person}{Daxin Jiang}.}
  \bibinfo{year}{2023}\natexlab{b}.
\newblock \showarticletitle{Wizardcoder: Empowering code large language models
  with evol-instruct}.
\newblock \bibinfo{journal}{\emph{arXiv preprint arXiv:2306.08568}}
  (\bibinfo{year}{2023}).
\newblock


\bibitem[McCabe(1977)]%
        {mc1977cycle}
\bibfield{author}{\bibinfo{person}{Thomas McCabe}.}
  \bibinfo{year}{1977}\natexlab{}.
\newblock \showarticletitle{A Complexity Measure}.
\newblock \bibinfo{journal}{\emph{Software Engineering, IEEE Transactions on}}
  \bibinfo{volume}{SE-2} (\bibinfo{date}{01} \bibinfo{year}{1977}),
  \bibinfo{pages}{308-- 320}.
\newblock
\urldef\tempurl%
\url{https://doi.org/10.1109/TSE.1976.233837}
\showDOI{\tempurl}


\bibitem[Mell et~al\mbox{.}(2024)]%
        {mell2024optimal}
\bibfield{author}{\bibinfo{person}{Stephen Mell}, \bibinfo{person}{Steve
  Zdancewic}, {and} \bibinfo{person}{Osbert Bastani}.}
  \bibinfo{year}{2024}\natexlab{}.
\newblock \showarticletitle{Optimal Program Synthesis via Abstract
  Interpretation}.
\newblock \bibinfo{journal}{\emph{Proceedings of the ACM on Programming
  Languages}} \bibinfo{volume}{8}, \bibinfo{number}{POPL}
  (\bibinfo{year}{2024}), \bibinfo{pages}{457--481}.
\newblock


\bibitem[Nakamaru et~al\mbox{.}(2020)]%
        {nakamaru2020empirical}
\bibfield{author}{\bibinfo{person}{Tomoki Nakamaru}, \bibinfo{person}{Tomomasa
  Matsunaga}, \bibinfo{person}{Tetsuro Yamazaki}, \bibinfo{person}{Soramichi
  Akiyama}, {and} \bibinfo{person}{Shigeru Chiba}.}
  \bibinfo{year}{2020}\natexlab{}.
\newblock \showarticletitle{An empirical study of method chaining in java}. In
  \bibinfo{booktitle}{\emph{Proceedings of the 17th International Conference on
  Mining Software Repositories}}. \bibinfo{pages}{93--102}.
\newblock


\bibitem[Nguyen et~al\mbox{.}(2024)]%
        {nguyen2024beginning}
\bibfield{author}{\bibinfo{person}{Sydney Nguyen},
  \bibinfo{person}{Hannah~McLean Babe}, \bibinfo{person}{Yangtian Zi},
  \bibinfo{person}{Arjun Guha}, \bibinfo{person}{Carolyn~Jane Anderson}, {and}
  \bibinfo{person}{Molly~Q Feldman}.} \bibinfo{year}{2024}\natexlab{}.
\newblock \showarticletitle{How Beginning Programmers and Code LLMs (Mis) read
  Each Other}. In \bibinfo{booktitle}{\emph{Proceedings of the CHI Conference
  on Human Factors in Computing Systems}}. \bibinfo{pages}{1--26}.
\newblock


\bibitem[{OpenAI}(2023a)]%
        {Codex}
\bibfield{author}{\bibinfo{person}{{OpenAI}}.}
  \bibinfo{year}{2023}\natexlab{a}.
\newblock \bibinfo{title}{{Codex}}.
\newblock
\newblock
\urldef\tempurl%
\url{https://openai.com/blog/openai-codex/}
\showURL{%
\tempurl}


\bibitem[{OpenAI}(2023b)]%
        {gpt4}
\bibfield{author}{\bibinfo{person}{{OpenAI}}.}
  \bibinfo{year}{2023}\natexlab{b}.
\newblock \bibinfo{title}{{gpt4}}.
\newblock
\newblock
\urldef\tempurl%
\url{https://cdn.openai.com/papers/gpt-4-system-card.pdf}
\showURL{%
\tempurl}


\bibitem[Ouyang et~al\mbox{.}(2022)]%
        {ouyang2022training}
\bibfield{author}{\bibinfo{person}{Long Ouyang}, \bibinfo{person}{Jeffrey Wu},
  \bibinfo{person}{Xu Jiang}, \bibinfo{person}{Diogo Almeida},
  \bibinfo{person}{Carroll Wainwright}, \bibinfo{person}{Pamela Mishkin},
  \bibinfo{person}{Chong Zhang}, \bibinfo{person}{Sandhini Agarwal},
  \bibinfo{person}{Katarina Slama}, \bibinfo{person}{Alex Ray},
  {et~al\mbox{.}}} \bibinfo{year}{2022}\natexlab{}.
\newblock \showarticletitle{Training language models to follow instructions
  with human feedback}.
\newblock \bibinfo{journal}{\emph{Advances in neural information processing
  systems}}  \bibinfo{volume}{35} (\bibinfo{year}{2022}),
  \bibinfo{pages}{27730--27744}.
\newblock


\bibitem[Pacheco and Ernst(2007)]%
        {pacheco2007randoop}
\bibfield{author}{\bibinfo{person}{Carlos Pacheco} {and}
  \bibinfo{person}{Michael~D Ernst}.} \bibinfo{year}{2007}\natexlab{}.
\newblock \showarticletitle{Randoop: feedback-directed random testing for
  Java}. In \bibinfo{booktitle}{\emph{Companion to the 22nd ACM SIGPLAN
  conference on Object-oriented programming systems and applications
  companion}}. \bibinfo{pages}{815--816}.
\newblock


\bibitem[{Pecan}(2024)]%
        {Pecan}
\bibfield{author}{\bibinfo{person}{{Pecan}}.} \bibinfo{year}{2024}\natexlab{}.
\newblock \bibinfo{title}{{Pecan GenAI Business}}.
\newblock
\newblock
\urldef\tempurl%
\url{https://www.pecan.ai/}
\showURL{%
\tempurl}


\bibitem[Peng et~al\mbox{.}(2023)]%
        {peng2023instruction}
\bibfield{author}{\bibinfo{person}{Baolin Peng}, \bibinfo{person}{Chunyuan Li},
  \bibinfo{person}{Pengcheng He}, \bibinfo{person}{Michel Galley}, {and}
  \bibinfo{person}{Jianfeng Gao}.} \bibinfo{year}{2023}\natexlab{}.
\newblock \showarticletitle{Instruction Tuning with GPT-4}.
\newblock \bibinfo{journal}{\emph{arXiv preprint arXiv:2304.03277}}
  (\bibinfo{year}{2023}).
\newblock


\bibitem[Peng et~al\mbox{.}(1997)]%
        {peng1997validity}
\bibfield{author}{\bibinfo{person}{Kaiping Peng}, \bibinfo{person}{Richard~E
  Nisbett}, {and} \bibinfo{person}{Nancy~YC Wong}.}
  \bibinfo{year}{1997}\natexlab{}.
\newblock \showarticletitle{Validity problems comparing values across cultures
  and possible solutions.}
\newblock \bibinfo{journal}{\emph{Psychological methods}} \bibinfo{volume}{2},
  \bibinfo{number}{4} (\bibinfo{year}{1997}), \bibinfo{pages}{329}.
\newblock


\bibitem[Perelman et~al\mbox{.}(2012)]%
        {partial-expressions}
\bibfield{author}{\bibinfo{person}{Daniel Perelman}, \bibinfo{person}{Sumit
  Gulwani}, \bibinfo{person}{Thomas Ball}, {and} \bibinfo{person}{Dan
  Grossman}.} \bibinfo{year}{2012}\natexlab{}.
\newblock \showarticletitle{Type-Directed Completion of Partial Expressions}.
  In \bibinfo{booktitle}{\emph{Proceedings of the 33rd ACM SIGPLAN Conference
  on Programming Language Design and Implementation}} (Beijing, China)
  \emph{(\bibinfo{series}{PLDI '12})}. \bibinfo{publisher}{Association for
  Computing Machinery}, \bibinfo{address}{New York, NY, USA},
  \bibinfo{pages}{275–286}.
\newblock
\showISBNx{9781450312059}
\urldef\tempurl%
\url{https://doi.org/10.1145/2254064.2254098}
\showDOI{\tempurl}


\bibitem[Polikarpova et~al\mbox{.}(2016)]%
        {polikarpova2016program}
\bibfield{author}{\bibinfo{person}{Nadia Polikarpova}, \bibinfo{person}{Ivan
  Kuraj}, {and} \bibinfo{person}{Armando Solar-Lezama}.}
  \bibinfo{year}{2016}\natexlab{}.
\newblock \showarticletitle{Program synthesis from polymorphic refinement
  types}.
\newblock \bibinfo{journal}{\emph{ACM SIGPLAN Notices}} \bibinfo{volume}{51},
  \bibinfo{number}{6} (\bibinfo{year}{2016}), \bibinfo{pages}{522--538}.
\newblock


\bibitem[Radford et~al\mbox{.}(2018)]%
        {radford2018improving}
\bibfield{author}{\bibinfo{person}{Alec Radford}, \bibinfo{person}{Karthik
  Narasimhan}, \bibinfo{person}{Tim Salimans}, \bibinfo{person}{Ilya
  Sutskever}, {et~al\mbox{.}}} \bibinfo{year}{2018}\natexlab{}.
\newblock \showarticletitle{Improving language understanding by generative
  pre-training}.
\newblock  (\bibinfo{year}{2018}).
\newblock


\bibitem[{Radon Team}(2024)]%
        {radon}
\bibfield{author}{\bibinfo{person}{{Radon Team}}.}
  \bibinfo{year}{2024}\natexlab{}.
\newblock \bibinfo{title}{{Radon's documentation.}}
\newblock
\newblock
\urldef\tempurl%
\url{https://radon.readthedocs.io/}
\showURL{%
\tempurl}


\bibitem[Rafailov et~al\mbox{.}(2024)]%
        {rafailov2024direct}
\bibfield{author}{\bibinfo{person}{Rafael Rafailov}, \bibinfo{person}{Archit
  Sharma}, \bibinfo{person}{Eric Mitchell}, \bibinfo{person}{Christopher~D
  Manning}, \bibinfo{person}{Stefano Ermon}, {and} \bibinfo{person}{Chelsea
  Finn}.} \bibinfo{year}{2024}\natexlab{}.
\newblock \showarticletitle{Direct preference optimization: Your language model
  is secretly a reward model}.
\newblock \bibinfo{journal}{\emph{Advances in Neural Information Processing
  Systems}}  \bibinfo{volume}{36} (\bibinfo{year}{2024}).
\newblock


\bibitem[Rajbhandari et~al\mbox{.}(2020)]%
        {rajbhandari2020zero}
\bibfield{author}{\bibinfo{person}{Samyam Rajbhandari}, \bibinfo{person}{Jeff
  Rasley}, \bibinfo{person}{Olatunji Ruwase}, {and} \bibinfo{person}{Yuxiong
  He}.} \bibinfo{year}{2020}\natexlab{}.
\newblock \showarticletitle{Zero: Memory optimizations toward training trillion
  parameter models}. In \bibinfo{booktitle}{\emph{SC20: International
  Conference for High Performance Computing, Networking, Storage and
  Analysis}}. IEEE, \bibinfo{pages}{1--16}.
\newblock


\bibitem[Rasley et~al\mbox{.}(2020)]%
        {rasley2020deepspeed}
\bibfield{author}{\bibinfo{person}{Jeff Rasley}, \bibinfo{person}{Samyam
  Rajbhandari}, \bibinfo{person}{Olatunji Ruwase}, {and}
  \bibinfo{person}{Yuxiong He}.} \bibinfo{year}{2020}\natexlab{}.
\newblock \showarticletitle{Deepspeed: System optimizations enable training
  deep learning models with over 100 billion parameters}. In
  \bibinfo{booktitle}{\emph{Proceedings of the 26th ACM SIGKDD International
  Conference on Knowledge Discovery \& Data Mining}}.
  \bibinfo{pages}{3505--3506}.
\newblock


\bibitem[Reimers and Gurevych(2019)]%
        {reimers2019sentence}
\bibfield{author}{\bibinfo{person}{Nils Reimers} {and} \bibinfo{person}{Iryna
  Gurevych}.} \bibinfo{year}{2019}\natexlab{}.
\newblock \showarticletitle{Sentence-bert: Sentence embeddings using siamese
  bert-networks}.
\newblock \bibinfo{journal}{\emph{arXiv preprint arXiv:1908.10084}}
  (\bibinfo{year}{2019}).
\newblock


\bibitem[Ren et~al\mbox{.}(2020)]%
        {ren2020codebleu}
\bibfield{author}{\bibinfo{person}{Shuo Ren}, \bibinfo{person}{Daya Guo},
  \bibinfo{person}{Shuai Lu}, \bibinfo{person}{Long Zhou},
  \bibinfo{person}{Shujie Liu}, \bibinfo{person}{Duyu Tang},
  \bibinfo{person}{Neel Sundaresan}, \bibinfo{person}{Ming Zhou},
  \bibinfo{person}{Ambrosio Blanco}, {and} \bibinfo{person}{Shuai Ma}.}
  \bibinfo{year}{2020}\natexlab{}.
\newblock \showarticletitle{Codebleu: a method for automatic evaluation of code
  synthesis}.
\newblock \bibinfo{journal}{\emph{arXiv preprint arXiv:2009.10297}}
  (\bibinfo{year}{2020}).
\newblock


\bibitem[Roziere et~al\mbox{.}(2023)]%
        {codellama}
\bibfield{author}{\bibinfo{person}{Baptiste Roziere}, \bibinfo{person}{Jonas
  Gehring}, \bibinfo{person}{Fabian Gloeckle}, \bibinfo{person}{Sten Sootla},
  \bibinfo{person}{Itai Gat}, \bibinfo{person}{Xiaoqing~Ellen Tan},
  \bibinfo{person}{Yossi Adi}, \bibinfo{person}{Jingyu Liu},
  \bibinfo{person}{Tal Remez}, \bibinfo{person}{J{\'e}r{\'e}my Rapin},
  {et~al\mbox{.}}} \bibinfo{year}{2023}\natexlab{}.
\newblock \showarticletitle{Code llama: Open foundation models for code}.
\newblock \bibinfo{journal}{\emph{arXiv preprint arXiv:2308.12950}}
  (\bibinfo{year}{2023}).
\newblock


\bibitem[Singhal et~al\mbox{.}(2023)]%
        {singhal2023large}
\bibfield{author}{\bibinfo{person}{Karan Singhal}, \bibinfo{person}{Shekoofeh
  Azizi}, \bibinfo{person}{Tao Tu}, \bibinfo{person}{S~Sara Mahdavi},
  \bibinfo{person}{Jason Wei}, \bibinfo{person}{Hyung~Won Chung},
  \bibinfo{person}{Nathan Scales}, \bibinfo{person}{Ajay Tanwani},
  \bibinfo{person}{Heather Cole-Lewis}, \bibinfo{person}{Stephen Pfohl},
  {et~al\mbox{.}}} \bibinfo{year}{2023}\natexlab{}.
\newblock \showarticletitle{Large language models encode clinical knowledge}.
\newblock \bibinfo{journal}{\emph{Nature}} \bibinfo{volume}{620},
  \bibinfo{number}{7972} (\bibinfo{year}{2023}), \bibinfo{pages}{172--180}.
\newblock


\bibitem[Solar-Lezama et~al\mbox{.}(2006)]%
        {solar2006combinatorial}
\bibfield{author}{\bibinfo{person}{Armando Solar-Lezama},
  \bibinfo{person}{Liviu Tancau}, \bibinfo{person}{Rastislav Bodik},
  \bibinfo{person}{Sanjit Seshia}, {and} \bibinfo{person}{Vijay Saraswat}.}
  \bibinfo{year}{2006}\natexlab{}.
\newblock \showarticletitle{Combinatorial sketching for finite programs}. In
  \bibinfo{booktitle}{\emph{Proceedings of the 12th international conference on
  Architectural support for programming languages and operating systems}}.
  \bibinfo{pages}{404--415}.
\newblock


\bibitem[Sotiropoulos et~al\mbox{.}(2024)]%
        {sotiropoulos2024api}
\bibfield{author}{\bibinfo{person}{Thodoris Sotiropoulos},
  \bibinfo{person}{Stefanos Chaliasos}, {and} \bibinfo{person}{Zhendong Su}.}
  \bibinfo{year}{2024}\natexlab{}.
\newblock \showarticletitle{API-Driven Program Synthesis for Testing Static
  Typing Implementations}.
\newblock  \bibinfo{volume}{8}, \bibinfo{number}{POPL}, Article
  \bibinfo{articleno}{62} (\bibinfo{date}{jan} \bibinfo{year}{2024}),
  \bibinfo{numpages}{32}~pages.
\newblock
\urldef\tempurl%
\url{https://doi.org/10.1145/3632904}
\showDOI{\tempurl}


\bibitem[{Stack Exchange Team}(2023)]%
        {stackexchange}
\bibfield{author}{\bibinfo{person}{{Stack Exchange Team}}.}
  \bibinfo{year}{2023}\natexlab{}.
\newblock \bibinfo{title}{{stackexchange QA communities.}}
\newblock
\newblock
\urldef\tempurl%
\url{https://huggingface.co/datasets/codefuse-ai/CodeExercise-Python-27k/}
\showURL{%
\tempurl}


\bibitem[Sun et~al\mbox{.}(2024)]%
        {sun2024neural}
\bibfield{author}{\bibinfo{person}{Zhensu Sun}, \bibinfo{person}{Xiaoning Du},
  \bibinfo{person}{Fu Song}, \bibinfo{person}{Shangwen Wang}, {and}
  \bibinfo{person}{Li Li}.} \bibinfo{year}{2024}\natexlab{}.
\newblock \showarticletitle{When Neural Code Completion Models Size up the
  Situation: Attaining Cheaper and Faster Completion through Dynamic Model
  Inference}. In \bibinfo{booktitle}{\emph{Proceedings of the IEEE/ACM 46th
  International Conference on Software Engineering}}. \bibinfo{pages}{1--12}.
\newblock


\bibitem[Thapa et~al\mbox{.}(2023)]%
        {thapa2023humans}
\bibfield{author}{\bibinfo{person}{Surendrabikram Thapa},
  \bibinfo{person}{Usman Naseem}, {and} \bibinfo{person}{Mehwish Nasim}.}
  \bibinfo{year}{2023}\natexlab{}.
\newblock \showarticletitle{From humans to machines: can ChatGPT-like LLMs
  effectively replace human annotators in NLP tasks}. In
  \bibinfo{booktitle}{\emph{Workshop Proceedings of the 17th International AAAI
  Conference on Web and Social Media}}.
\newblock


\bibitem[Ullah et~al\mbox{.}(2024)]%
        {ullah2024llms}
\bibfield{author}{\bibinfo{person}{Saad Ullah}, \bibinfo{person}{Mingji Han},
  \bibinfo{person}{Saurabh Pujar}, \bibinfo{person}{Hammond Pearce},
  \bibinfo{person}{Ayse Coskun}, {and} \bibinfo{person}{Gianluca Stringhini}.}
  \bibinfo{year}{2024}\natexlab{}.
\newblock \showarticletitle{LLMs Cannot Reliably Identify and Reason About
  Security Vulnerabilities (Yet?): A Comprehensive Evaluation, Framework, and
  Benchmarks}. In \bibinfo{booktitle}{\emph{IEEE Symposium on Security and
  Privacy}}.
\newblock


\bibitem[Van~der Maaten and Hinton(2008)]%
        {van2008visualizing}
\bibfield{author}{\bibinfo{person}{Laurens Van~der Maaten} {and}
  \bibinfo{person}{Geoffrey Hinton}.} \bibinfo{year}{2008}\natexlab{}.
\newblock \showarticletitle{Visualizing data using t-SNE.}
\newblock \bibinfo{journal}{\emph{Journal of machine learning research}}
  \bibinfo{volume}{9}, \bibinfo{number}{11} (\bibinfo{year}{2008}).
\newblock


\bibitem[Vaswani et~al\mbox{.}(2017)]%
        {vaswani2017attention}
\bibfield{author}{\bibinfo{person}{Ashish Vaswani}, \bibinfo{person}{Noam
  Shazeer}, \bibinfo{person}{Niki Parmar}, \bibinfo{person}{Jakob Uszkoreit},
  \bibinfo{person}{Llion Jones}, \bibinfo{person}{Aidan~N Gomez},
  \bibinfo{person}{{\L}ukasz Kaiser}, {and} \bibinfo{person}{Illia
  Polosukhin}.} \bibinfo{year}{2017}\natexlab{}.
\newblock \showarticletitle{Attention is all you need}.
\newblock \bibinfo{journal}{\emph{Advances in neural information processing
  systems}}  \bibinfo{volume}{30} (\bibinfo{year}{2017}).
\newblock


\bibitem[Wang et~al\mbox{.}(2024a)]%
        {wang2024exploring}
\bibfield{author}{\bibinfo{person}{Chaozheng Wang}, \bibinfo{person}{Zongjie
  Li}, \bibinfo{person}{Cuiyun Gao}, \bibinfo{person}{Wenxuan Wang},
  \bibinfo{person}{Ting Peng}, \bibinfo{person}{Hailiang Huang},
  \bibinfo{person}{Yuetang Deng}, \bibinfo{person}{Shuai Wang}, {and}
  \bibinfo{person}{Michael~R Lyu}.} \bibinfo{year}{2024}\natexlab{a}.
\newblock \showarticletitle{Exploring Multi-Lingual Bias of Large Code Models
  in Code Generation}.
\newblock \bibinfo{journal}{\emph{arXiv preprint arXiv:2404.19368}}
  (\bibinfo{year}{2024}).
\newblock


\bibitem[Wang et~al\mbox{.}(2024b)]%
        {wang2024diversity}
\bibfield{author}{\bibinfo{person}{Peiqi Wang}, \bibinfo{person}{Yikang Shen},
  \bibinfo{person}{Zhen Guo}, \bibinfo{person}{Matthew Stallone},
  \bibinfo{person}{Yoon Kim}, \bibinfo{person}{Polina Golland}, {and}
  \bibinfo{person}{Rameswar Panda}.} \bibinfo{year}{2024}\natexlab{b}.
\newblock \showarticletitle{Diversity Measurement and Subset Selection for
  Instruction Tuning Datasets}.
\newblock \bibinfo{journal}{\emph{arXiv preprint arXiv:2402.02318}}
  (\bibinfo{year}{2024}).
\newblock


\bibitem[Wang et~al\mbox{.}(2017)]%
        {wang2017program}
\bibfield{author}{\bibinfo{person}{Xinyu Wang}, \bibinfo{person}{Isil Dillig},
  {and} \bibinfo{person}{Rishabh Singh}.} \bibinfo{year}{2017}\natexlab{}.
\newblock \showarticletitle{Program synthesis using abstraction refinement}.
\newblock \bibinfo{journal}{\emph{Proceedings of the ACM on Programming
  Languages}} \bibinfo{volume}{2}, \bibinfo{number}{POPL}
  (\bibinfo{year}{2017}), \bibinfo{pages}{1--30}.
\newblock


\bibitem[Wang et~al\mbox{.}(2023)]%
        {wang2023self}
\bibfield{author}{\bibinfo{person}{Yizhong Wang}, \bibinfo{person}{Yeganeh
  Kordi}, \bibinfo{person}{Swaroop Mishra}, \bibinfo{person}{Alisa Liu},
  \bibinfo{person}{Noah~A Smith}, \bibinfo{person}{Daniel Khashabi}, {and}
  \bibinfo{person}{Hannaneh Hajishirzi}.} \bibinfo{year}{2023}\natexlab{}.
\newblock \showarticletitle{Self-Instruct: Aligning Language Models with
  Self-Generated Instructions}. In \bibinfo{booktitle}{\emph{Proceedings of the
  61st Annual Meeting of the Association for Computational Linguistics (Volume
  1: Long Papers)}}. \bibinfo{pages}{13484--13508}.
\newblock


\bibitem[Wang et~al\mbox{.}(2021)]%
        {wang2021codet5}
\bibfield{author}{\bibinfo{person}{Yue Wang}, \bibinfo{person}{Weishi Wang},
  \bibinfo{person}{Shafiq~R. Joty}, {and} \bibinfo{person}{Steven C.~H. Hoi}.}
  \bibinfo{year}{2021}\natexlab{}.
\newblock \showarticletitle{CodeT5: Identifier-aware Unified Pre-trained
  Encoder-Decoder Models for Code Understanding and Generation}. In
  \bibinfo{booktitle}{\emph{Proceedings of the 2021 Conference on Empirical
  Methods in Natural Language Processing, {EMNLP} 2021, Virtual Event / Punta
  Cana, Dominican Republic, 7-11 November, 2021}},
  \bibfield{editor}{\bibinfo{person}{Marie{-}Francine Moens},
  \bibinfo{person}{Xuanjing Huang}, \bibinfo{person}{Lucia Specia}, {and}
  \bibinfo{person}{Scott~Wen{-}tau Yih}} (Eds.).
  \bibinfo{publisher}{Association for Computational Linguistics}.
\newblock


\bibitem[Wei et~al\mbox{.}(2021)]%
        {wei2021finetuned}
\bibfield{author}{\bibinfo{person}{Jason Wei}, \bibinfo{person}{Maarten Bosma},
  \bibinfo{person}{Vincent~Y Zhao}, \bibinfo{person}{Kelvin Guu},
  \bibinfo{person}{Adams~Wei Yu}, \bibinfo{person}{Brian Lester},
  \bibinfo{person}{Nan Du}, \bibinfo{person}{Andrew~M Dai}, {and}
  \bibinfo{person}{Quoc~V Le}.} \bibinfo{year}{2021}\natexlab{}.
\newblock \showarticletitle{Finetuned language models are zero-shot learners}.
\newblock \bibinfo{journal}{\emph{arXiv preprint arXiv:2109.01652}}
  (\bibinfo{year}{2021}).
\newblock


\bibitem[Wei et~al\mbox{.}({[n.\,d.]})]%
        {weimagicoder}
\bibfield{author}{\bibinfo{person}{Yuxiang Wei}, \bibinfo{person}{Zhe Wang},
  \bibinfo{person}{Jiawei Liu}, \bibinfo{person}{Yifeng Ding}, {and}
  \bibinfo{person}{Lingming Zhang}.} \bibinfo{year}{[n.\,d.]}\natexlab{}.
\newblock \showarticletitle{Magicoder: Empowering Code Generation with
  OSS-Instruct}. In \bibinfo{booktitle}{\emph{Forty-first International
  Conference on Machine Learning}}.
\newblock


\bibitem[Wei et~al\mbox{.}(2023)]%
        {wei2023magicoder}
\bibfield{author}{\bibinfo{person}{Yuxiang Wei}, \bibinfo{person}{Zhe Wang},
  \bibinfo{person}{Jiawei Liu}, \bibinfo{person}{Yifeng Ding}, {and}
  \bibinfo{person}{Lingming Zhang}.} \bibinfo{year}{2023}\natexlab{}.
\newblock \showarticletitle{Magicoder: Source code is all you need}.
\newblock \bibinfo{journal}{\emph{arXiv preprint arXiv:2312.02120}}
  (\bibinfo{year}{2023}).
\newblock


\bibitem[Xu et~al\mbox{.}(2023)]%
        {xu2023wizardlm}
\bibfield{author}{\bibinfo{person}{Can Xu}, \bibinfo{person}{Qingfeng Sun},
  \bibinfo{person}{Kai Zheng}, \bibinfo{person}{Xiubo Geng},
  \bibinfo{person}{Pu Zhao}, \bibinfo{person}{Jiazhan Feng},
  \bibinfo{person}{Chongyang Tao}, {and} \bibinfo{person}{Daxin Jiang}.}
  \bibinfo{year}{2023}\natexlab{}.
\newblock \showarticletitle{Wizardlm: Empowering large language models to
  follow complex instructions}.
\newblock \bibinfo{journal}{\emph{arXiv preprint arXiv:2304.12244}}
  (\bibinfo{year}{2023}).
\newblock


\bibitem[Yoo and Harman(2007)]%
        {yoo2007pareto}
\bibfield{author}{\bibinfo{person}{Shin Yoo} {and} \bibinfo{person}{Mark
  Harman}.} \bibinfo{year}{2007}\natexlab{}.
\newblock \showarticletitle{Pareto efficient multi-objective test case
  selection}. In \bibinfo{booktitle}{\emph{Proceedings of the 2007
  international symposium on Software testing and analysis}}.
  \bibinfo{pages}{140--150}.
\newblock


\bibitem[Zhang et~al\mbox{.}(2024a)]%
        {zhang2024scaling}
\bibfield{author}{\bibinfo{person}{Biao Zhang}, \bibinfo{person}{Zhongtao Liu},
  \bibinfo{person}{Colin Cherry}, {and} \bibinfo{person}{Orhan Firat}.}
  \bibinfo{year}{2024}\natexlab{a}.
\newblock \showarticletitle{When Scaling Meets LLM Finetuning: The Effect of
  Data, Model and Finetuning Method}.
\newblock \bibinfo{journal}{\emph{arXiv preprint arXiv:2402.17193}}
  (\bibinfo{year}{2024}).
\newblock


\bibitem[Zhang et~al\mbox{.}(2024b)]%
        {zhang2024large}
\bibfield{author}{\bibinfo{person}{Zhuo Zhang}, \bibinfo{person}{Guangyu Shen},
  \bibinfo{person}{Guanhong Tao}, \bibinfo{person}{Siyuan Cheng}, {and}
  \bibinfo{person}{Xiangyu Zhang}.} \bibinfo{year}{2024}\natexlab{b}.
\newblock \showarticletitle{On large language models’ resilience to coercive
  interrogation}. In \bibinfo{booktitle}{\emph{2024 IEEE Symposium on Security
  and Privacy (SP)}}. IEEE Computer Society, \bibinfo{pages}{252--252}.
\newblock


\bibitem[Zheng et~al\mbox{.}(2023)]%
        {zheng2023lmsys}
\bibfield{author}{\bibinfo{person}{Lianmin Zheng}, \bibinfo{person}{Wei-Lin
  Chiang}, \bibinfo{person}{Ying Sheng}, \bibinfo{person}{Tianle Li},
  \bibinfo{person}{Siyuan Zhuang}, \bibinfo{person}{Zhanghao Wu},
  \bibinfo{person}{Yonghao Zhuang}, \bibinfo{person}{Zhuohan Li},
  \bibinfo{person}{Zi Lin}, \bibinfo{person}{Eric Xing}, {et~al\mbox{.}}}
  \bibinfo{year}{2023}\natexlab{}.
\newblock \showarticletitle{Lmsys-chat-1m: A large-scale real-world llm
  conversation dataset}.
\newblock \bibinfo{journal}{\emph{arXiv preprint arXiv:2309.11998}}
  (\bibinfo{year}{2023}).
\newblock


\bibitem[Zheng et~al\mbox{.}(2024)]%
        {zheng2024judging}
\bibfield{author}{\bibinfo{person}{Lianmin Zheng}, \bibinfo{person}{Wei-Lin
  Chiang}, \bibinfo{person}{Ying Sheng}, \bibinfo{person}{Siyuan Zhuang},
  \bibinfo{person}{Zhanghao Wu}, \bibinfo{person}{Yonghao Zhuang},
  \bibinfo{person}{Zi Lin}, \bibinfo{person}{Zhuohan Li},
  \bibinfo{person}{Dacheng Li}, \bibinfo{person}{Eric Xing}, {et~al\mbox{.}}}
  \bibinfo{year}{2024}\natexlab{}.
\newblock \showarticletitle{Judging llm-as-a-judge with mt-bench and chatbot
  arena}.
\newblock \bibinfo{journal}{\emph{Advances in Neural Information Processing
  Systems}}  \bibinfo{volume}{36} (\bibinfo{year}{2024}).
\newblock


\bibitem[Zhou et~al\mbox{.}(2024)]%
        {zhou2024lima}
\bibfield{author}{\bibinfo{person}{Chunting Zhou}, \bibinfo{person}{Pengfei
  Liu}, \bibinfo{person}{Puxin Xu}, \bibinfo{person}{Srinivasan Iyer},
  \bibinfo{person}{Jiao Sun}, \bibinfo{person}{Yuning Mao},
  \bibinfo{person}{Xuezhe Ma}, \bibinfo{person}{Avia Efrat},
  \bibinfo{person}{Ping Yu}, \bibinfo{person}{Lili Yu}, {et~al\mbox{.}}}
  \bibinfo{year}{2024}\natexlab{}.
\newblock \showarticletitle{Lima: Less is more for alignment}.
\newblock \bibinfo{journal}{\emph{Advances in Neural Information Processing
  Systems}}  \bibinfo{volume}{36} (\bibinfo{year}{2024}).
\newblock


\end{thebibliography}

\end{document}